
\input harvmac  
\input epsf
\noblackbox
\figno=0
\def\fig#1#2#3{
\par\begingroup\parindent=0pt\leftskip=1cm\rightskip=1cm\parindent=0pt
\baselineskip=11pt
\global\advance\figno by 1
\midinsert
\epsfxsize=#3
\centerline{\epsfbox{#2}}
\vskip 12pt
{\bf Fig. \the\figno:} #1\par
\endinsert\endgroup\par
}
\def\figlabel#1{\xdef#1{\the\figno}}
\def\encadremath#1{\vbox{\hrule\hbox{\vrule\kern8pt\vbox{\kern8pt
\hbox{$\displaystyle #1$}\kern8pt}
\kern8pt\vrule}\hrule}}

\def\encadremath#1{\vbox{\hrule\hbox{\vrule\kern8pt\vbox{\kern8pt
 \hbox{$\displaystyle #1$}\kern8pt}
 \kern8pt\vrule}\hrule}}


 \def\inbar{\vrule height1.5ex width.4pt depth0pt}
 \def\IC{\relax\,\hbox{$\inbar\kern-.3em{\rm C}$}}
 \def\IN{\relax{\rm I\kern-.18em N}}
 \def\IP{\relax{\rm I\kern-.18em P}}

  \def\IZ{\relax\ifmmode\mathchoice
 {\hbox{\cmss Z\kern-.4em Z}}{\hbox{\cmss Z\kern-.4em Z}}
 {\lower.9pt\hbox{\cmsss Z\kern-.4em Z}}
 {\lower1.2pt\hbox{\cmsss Z\kern-.4em Z}}\else{\cmss Z\kern-.4em 
Z}\fi}
 \def\IB{\relax{\rm I\kern-.18em B}}
 \def\IC{{\relax\hbox{$\inbar\kern-.3em{\rm C}$}}}
 \def\Ic{{\relax\hbox{$\inbar\kern-.22em{\rm c}$}}}
 \def\ID{\relax{\rm I\kern-.18em D}}
 \def\IE{\relax{\rm I\kern-.18em E}}
 \def\IF{\relax{\rm I\kern-.18em F}}
 \def\IG{\relax\hbox{$\inbar\kern-.3em{\rm G}$}}
 \def\IGa{\relax\hbox{${\rm I}\kern-.18em\Gamma$}}
 \def\IH{\relax{\rm I\kern-.18em H}}
 \def\II{\relax{\rm I\kern-.18em I}}
 \def\IK{\relax{\rm I\kern-.18em K}}
 \def\IP{\relax{\rm I\kern-.18em P}}
 \def\IR{\relax{\rm I\kern-.18em R}}
\font\cmss=cmss10
 \font\cmsss=cmss10 at 7pt
\font\litfont = cmr6


\def\eqnn#1{\xdef
#1{(\secsym\the\meqno)}\writedef{#1\leftbracket#1}%
 \global\advance\meqno by1\wrlabeL#1}
 \def\eqna#1{\xdef
#1##1{\hbox{$(\secsym\the\meqno##1)$}}

\writedef{#1\numbersign1\leftbracket#1{\numbersign1}}%
 \global\advance\meqno by1\wrlabeL{#1$\{\}$}}
 \def\eqn#1#2{\xdef
#1{(\secsym\the\meqno)}\writedef{#1\leftbracket#1}%
 \global\advance\meqno by1$$#2\eqno#1\eqlabeL#1$$}

\def\inv{^{-1}}
\def\Tr{{\rm Tr}}
\def\hf{\textstyle{1\over 2}}

 \def\frac#1#2{{\textstyle{#1\over#2}}}
\def\inv{^{\raise.15ex\hbox{${\scriptscriptstyle -}$}\kern-.05em 1}}
 
\def\[{\left[}
\def\]{\right]}
\def\({\left(}
\def\){\right)}
\def\<{\left\langle\,}
\def\>{\,\right\rangle}
 
 %

  \def\R{\relax{\rm I\kern-.18em R}}
  \font\cmss=cmss10 \font\cmsss=cmss10 at 7pt 
  \def\Z{\relax\ifmmode\mathchoice
  {\hbox{\cmss Z\kern-.4em Z}}{\hbox{\cmss Z\kern-.4em Z}} 
  {\lower.9pt\hbox{\cmsss Z\kern-.4em Z}}
  {\lower1.2pt\hbox{\cmsss Z\kern-.4em Z}}\else{\cmss Z\kern-.4em 
  Z}\fi} 
\def\bigone{\hbox{1\kern -.23em {\rm l}}}

\def\hf{{\litfont {1 \over 2}}}

\def\Im{{\rm Im ~}}

\def\p{\partial}


\def\b{\beta}
\def\g{\gamma}

\def\e{\epsilon}

\def\t{\tau}

\def\G{\Gamma}

\def\L{\Lambda}


   \def\Im{{\rm Im}}
    
  \def\CA {{\cal A}}
  \def\CB {{\cal B}}
  \def\CC {{\cal C}}

  \def\CH {{\cal H}}

  \def\CM {{\cal M}}

  \def\CP {{\cal P}}

  \def\CS {{\cal S}}

  \def\CV {{\cal V}}

  \def\CZ {{\cal Z}}
 %

\chardef\tempcat=\the\catcode`\@ \catcode`\@=11
\def\cyracc{\def\u##1{\if \i##1\accent"24 i%
    \else \accent"24 ##1\fi }}
\newfam\cyrfam
\font\tencyr=wncyr10
\def\cyr{\fam\cyrfam\tencyr\cyracc}




  \def\11{{\bigone}}
 



\def\np#1#2#3{{Nucl. Phys.} {\CB#1} (#2) #3}
\def\pl#1#2#3{{Phys. Lett. }{\CB#1} (#2) #3}

\def\hepth#1{{hep-th/}#1}

   \lref\KP{I. Kostov and V. Petkova, in preparation.}

 \lref\AlexandrovNN{ S.~Y.~Alexandrov, V.~A.~Kazakov and
D.~Kutasov, ``Non-Perturbative Effects in Matrix Models and
D-branes,''  hep-th/0306177.
}

\lref\DiFrancescoGinsparg{P.~Di Francesco, P.~Ginsparg and
J.~Zinn-Justin,``2-D Gravity and random matrices,'' Phys.\ Rept.\
{\bf 254} (1995) 1,  hep-th/9306153.
}

\lref\polchinski{J. Polchinski, ``What is string theory'',
{\it Lectures presented at the 1994 Les Houches Summer School
``Fluctuating Geometries in Statistical Mechanics and Field 
Theory''},  
\hepth{9411028}.}

\lref\KlebanovMQM{I. Klebanov, {\it Lectures delivered at the ICTP
Spring School on String Theory and Quantum Gravity},
Trieste, April 1991, \hepth{9108019}.}

\lref\SchomerusVV{ V.~Schomerus, ``Rolling tachyons from Liouville
theory,''  hep-th/0306026.
}

\lref\Icar{
I. Kostov , ``Solvable Statistical Models on Random Lattices",
Proceedings of the Conference on recent developments in statistical
mechanics and quantum field theory.  (Trieste, 10 - 12 April 1995),
Nucl. Phys. B (Proc. Suppl.) 45 A (1996) 13-28, hep-th/9509124 .
}

\lref\Ion{I. Kostov,  ``$O(n)$ vector model on a planar random 
 lattice: spectrum of anomalous dimensions", Mod.Phys.Lett.{ A4} (1989) 217.
}

\lref\McGreevyEP{
J.~McGreevy, J.~Teschner and H.~Verlinde,
``Classical and quantum D-branes in 2D string theory,''
hep-th/0305194.
}

\lref\TeschnerQK{ J.~Teschner, ``On boundary perturbations in
Liouville theory and brane dynamics in noncritical string
theories,''  hep-th/0308140.
}

\lref\DV{R. Dijkgraaf,  C. Vafa, 
``N=1 Supersymmetry, Deconstruction, and Bosonic Gauge Theories'',
hep-th/0302011.
}
 
  \lref\kkk{V. Kazakov,  I. Kostov,  D.  Kutasov,
  ``A Matrix Model for the Two Dimensional Black Hole",
  Nucl.Phys. B622 (2002) 141,
  hep-th/0101011.
  }

   \lref\JevickiQN{
A.~Jevicki,
``Developments in 2-d string theory,''
 hep-th/9309115.
}

\lref\Sennew{ A.~Sen, ``Open-Closed Duality: Lessons from the
Matrix Model,''  hep-th/0308068.
}
   \lref\Kstau{I. Kostov and 
   M. Staudacher, ``Strings in discrete and continuous 
target  spaces:
a comparison", \pl{305}{1993}{43}.
   }
   \lref\FZZb{V.~Fateev, A.~B.~Zamolodchikov and A.~B.~Zamolodchikov,
``Boundary Liouville field theory. I: Boundary state and boundary
two-point function,''  hep-th/0001012.
}

 \lref\DO{
H. Dorn, H.J. Otto: 
Two and three point functions in Liouville theory, 
\np{429}{1994}{375}, hep-th/9403141.
}
 
\lref\PTtwo{B.~Ponsot, J.~Teschner, ``Boundary Liouville Field 
Theory: Boundary three point function'',
  Nucl.~Phys.~{B622} (2002) 309, \hepth{0110244}.
  }

\lref\Teschner{
J. Teschner. On the Liouville Three-Point Function.
Phys.Lett., B363 (1995) 65. 
}
  \lref\ReSch{A. Recknagel,  V. Schomerus,
  ``Boundary Deformation Theory and Moduli Spaces of D-Branes",
   Nucl.Phys. B545 (1999) 233, hep-th/9811237.
  }  

  \lref\DiK{
  P. Di Francesco,  D. Kutasov, 
  ``World Sheet and Space Time Physics in
   Two Dimensional (Super) String   Theory", Nucl.Phys. B375 (1992) 
119,   hep-th/9109005.
}
  \lref\KMSnew{
I. R. Klebanov,  J. Maldacena,  N. Seiberg,
``Unitary and Complex Matrix Models as 1-d Type 0 Strings'',
   hep-th/0309168.
   }
  \lref\GM{
  P. Ginsparg and G. Moore,
  ``Lectures on 2D gravity and 2D string theory (TASI 1992)", 
  hep-th/9304011.
   }

\lref\Gouli{
M. Goulian and B. Li, Phys. Rev. Lett. 66 (1991), 2051.
}
 
\lref\VDotsenko{
V. Dotsenko, Mod. Phys. Lett. A6(1991), 3601.
}

 \lref\hosomichi{K.~Hosomichi, "Bulk-Boundary Propagator in 
Liouville 
Theory on a Disc", JHEP 0111 
 044 (2001), \hepth{0108093}.
 }

  \lref\Witten{
  E.~Witten,
``Ground ring of two-dimensional string theory,''
Nucl.\ Phys.\ B { 373}, 187 (1992),
hep-th/9108004.
}

\lref\WitZw{ E.~Witten and B.~Zwiebach, ``Algebraic structures
and differential geometry in 2D string theory,'' Nucl.\ Phys.\
B { 377}, 55 (1992),  hep-th/9201056.
}

 \lref\KMS{ D. Kutasov, E. Martinec, N. Seiberg, 
   ``Ground rings and their modules in 2-D gravity with $c\le1 $ 
matter",  
      Phys.Lett. {\CB276} (1992) 437, \hepth{9111048}.
}

 \lref\KlebanovMQM{
  I.~Klebanov, ``String theory in two-dimensions,''
 hep-th/9108019.
}
  \lref\bershkut{
  M. Bershadsky and D. Kutasov,
 ``Scattering of open and closed strings in (1+1)-dimensions",
 \np{382}{1992}{213}, \hepth{9204049}.
  }
  \lref\berkut{
  M. Bershadsky and D. Kutasov,
  Phys. Lett. 274B (1992) 331,  hep-th/9110034.
  }
  \lref\KKK{V. Kazakov,  I. Kostov,  D.  Kutasov,
  ``A Matrix Model for the Two Dimensional Black Hole",
  Nucl.Phys. B622 (2002) 141,
  hep-th/0101011.
  }

\lref\Isixv{I. Kostov, ``Exact solution of the six-Vertex Model 
 on a random lattice",
hep-th/9907060.}

\lref\Ise{I. Kostov, ``String Equation for String Theory on a 
Circle'',
\np{624}{2002}{146}, \hepth{0107247}.
}

 \lref\Idis{I. Kostov, ``Strings with discrete target space'',
 \np{376}{1992}{539},  hep-th/9112059.
}

\lref\KKloop{
V. Kazakov, I. Kostov,
``Loop Gas Model for Open Strings'',
 \np{386}{1992}{520}.
}
 \lref\XID{Xi Yin, 
 ``Matrix Models, Integrable Structures, and T-duality of Type 0 
String   Theory",
 hep-th/0312236.
 }  
\lref\MukhiImbimbo{ C. Imbimbo, S. Mukhi, \np{449}{1995}{553}, 
hep-th/9505127.}
  \lref\mukhi{S. Mukhi, ``Topological Matrix Models, Liouville Matrix 
Model and c=1 String Theory'',
  hep-th/0310287.}

\lref\ZZtp{A.B. Zamolodchikov, Al.B. Zamolodchikov: 
Structure constants and conformal bootstrap in Liouville field theory,
\np{477}{1996}{577}, hep-th/9506136.
}

\lref\Ibliou{
I.  Kostov, ``Boundary Correlators in 2D Quantum 
Gravity:
 Liouville versus Discrete Approach ", \np{658}{2003}{397},
 hep-th/0212194.}
 \lref\KPS{I. Kostov, B. Ponsot and D. Serban,   ``Boundary  
Liouville Theory and 
2D Quantum Gravity", hep-th/0307189.}

\lref\ZZPseudo{
A.~B. Zamolodchikov and A.~B. Zamolodchikov, ``Liouville field theory 
on a
  pseudosphere,  
  \hepth{0101152}.
}

\lref\GovJcf{
S. Govindarajan, T. Jayaraman and V. John,
ÒGenus Zero Correlation Functions in
 $c<1$ String Theory,Ó Phys. Rev. D 48 (1993) 839,
\hepth{9208064}.
}

\lref\MartinecKA{ E.~J.~Martinec, ``The annular report on 
non-critical string
theory,''  hep-th/0305148.}

  \lref\GovLast{
  S. Govindarajan, T. Jayaraman and V. John,
  ``Correlation Functions and Multicritical Flows in $c<1$ String 
Theory",
 Int. J. Mod.Phys. A10 (1995) 477,  hep-th/9309040.}
     
   \lref\TY{Y. Tanii, S.-I. Yamaguchi, ``Two-dimensional 
   quantum gravity on a disc", Mod. Phys. Lett. A7 (1992) 521,
 hep-th/9110068; ``Disk Amplitudes in Two-Dimensional Open String 
Theories", hep-th/9203002.
  }
 \lref\ValyaJB{
 V.B. Petkova,  J.-B. Zuber, ``BCFT: from the boundary to the bulk'',
 Talk presented at TMR-conference "Nonperturbative Quantum Effects   
2000", 
hep-th/0009219.
  }
 \lref\higkos{
 S. Higuchi and  I. Kostov , ``Feynman rules for   
string 
theories with 
discrete target space", \pl{357}{1995}{62}, hep-th/9506022. }
 \lref\topint{M. Aganagic,  R. Dijkgraaf,  A. Klemm,  M. Marino,  C. 
Vafa,
 ``Topological Strings and Integrable Hierarchies'',
 hep-th/0312085.
 }

 \lref\Iconfm{ I. Kostov, 
 ``Conformal Field Theory Techniques in Random Matrix models", 
  hep-th/9907060.}

 \lref\ADEold{
I. Kostov, ``The ADE face models on a fluctuating planar lattice",
\np{326}{1989}{583}.
}

\lref\newhat{
M.~R.~Douglas, I.~R.~Klebanov, D.~Kutasov, J.~Maldacena, E.~Martinec 
and N.~Seiberg,
``A new hat for the c = 1 matrix model,'' hep-th/0307195.}

   \lref\KlebPol{
I.~R.~Klebanov and A.~M.~Polyakov, ``Interaction of discrete
states in two-dimensional string theory,'' Mod.\ Phys.\ Lett.\ A
{6}, 3273 (1991) hep-th/9109032.
}

   \lref\DMP{ 
R.~Dijkgraaf, G.~W.~Moore and R.~Plesser,
``The Partition function of 2-D string theory,''
Nucl. Phys. B {394}, 356 (1993),
 hep-th/9208031.}
 
\lref\AKK{
S.  Alexandrov,  V. Kazakov,  I.  Kostov, 
``Time-dependent backgrounds of 2D string theory",
  Nucl.Phys. B640 (2002) 119,  hep-th/0205079.
  }
  \lref\Iflows{I. Kostov, ``Integrable flows in $c=1$ string theory",
  J.Phys. A36 (2003) 3153,  hep-th/0208034.}
  
  \lref\adem{ I. Kostov, 
  ``Gauge Invariant Matrix Model for the \^A-\^D-\^E Closed Strings",
  Phys.Lett. B297 (1992) 74-81, hep-th/9208053.
  }

   \lref\ADKMV{
  M.~Aganagic, R.~Dijkgraaf, A.~Klemm, M.~Marino and C.~Vafa,
``Topological Strings and Integrable Hierarchies,''  hep-th/0312085.
  }
 \lref\SeibergS{
 N. Seiberg and D. Shih, ``Branes, rings and matrix models on minimal 
(super)string theory",  hep-th/0312170.
 }

\lref\Iham{
I. Kostov,
``Loop space Hamiltonian for $c \le 1$ open strings'',
Phys.Lett. B349 (1995) 284, hep-th/9501135.
}

\lref\Iopen{
I. Kostov,
``Field Theory of Open and Closed Strings with Discrete Target 
Space'',
Phys.Lett. B344 (1995) 135,
    hep-th/9410164.}
\lref\mgv{ J.~McGreevy and H.~Verlinde, ``Strings from tachyons'' ,
JHEP 0312 (2003) 054, 
 hep-th/0304224.
}

\lref\KlebanovKM{
I.~R.~Klebanov, J.~Maldacena and N.~Seiberg,
``D-brane decay in two-dimensional string theory,''
hep-th/0305159.
}
 
\overfullrule=0pt
\Title{\vbox{\baselineskip12pt\hbox
{SPhT-T03/214}\hbox{hep-th/0312301 }}}
{\vbox{\centerline
 {Boundary Ground Ring  in 2D  String Theory }
 \centerline{}
\centerline{ }
 \vskip2pt
}}
 %
\centerline{Ivan K. Kostov\footnote{$^\ast$}{{\tt 
kostov@spht.saclay.cea.fr}}\footnote{$^{\dag}$}{{ 
Associate member of {\cyr  IYaIYaE -- BAN},  \ 
Sofia, Bulgaria }}}

\centerline{{\it  Service de Physique 
Th{\'e}orique,
 CNRS -- URA 2306, }}
\centerline{{\it C.E.A. - Saclay,   
  F-91191 Gif-Sur-Yvette, France\footnote{$^\#$}{Permanent address}}}

 \lref\kachru{S. Kachru, 
 ``Quantum Rings and Recursion Relations in 2D Quantum Gravity'',
 Mod. Phys. Lett. A7 (1992) 1419, 
  hep-th/9201072.}

\lref\BPZ{A. Belavin, A. Polyakov, A. Zamolodchikov,
  ``Infinite 
 conformal symmetry in two-dimensional quantum 
field theory",
 Nucl. Phys. {\CB241}, 333 (1984).
}

  \lref\CMM{S. Kharchev,  A. Marshakov,  A. Mironov,  A. Morozov,  S. 
Pakuliak,
  ``Conformal Matrix Models as an Alternative to Conventional 
Multi-Matrix   Models'', Nucl. Phys. B404 (1993) 717, 
   hep-th/9208044.
   }

 
\vskip 2cm

{\ninepoint

\noindent{  
  The  2D quantum gravity  on a disc, or the non-critical theory of  
open strings,
  is known to exhibit an  integrable structure,  the boundary ground 
ring, which 
  determines completely the  boundary correlation functions.
 Inspired by the recent progress   in boundary Liouville theory,  we  
extend the  
  ground ring  relations to the case of  non-vanishing boundary 
Liouville
 interaction  known also as FZZT brane in the context of the 2D 
string theory.  
The ring  relations  yield an  over-determined set of functional  
recurrence equations  for the boundary correlation functions.  
The ring  action closes on an infinite array of equally spaced FZZT 
branes
for which we propose  a matrix model realization.  In this matrix 
model the
boundary ground ring  is generated by a  pair of  complex matrix 
fields.
   }
  }

\Date{  December 2003}
\vfill
\eject



\baselineskip=14pt plus 1pt minus 1pt

\newsec{Introduction}
  
  \def\QG{2D QG}
  
\noindent
The  world-sheet description of the non-critical string theories is 
given by the
two-dimensional quantum gravity (for reviews see
\refs{\GM\DiFrancescoGinsparg\polchinski\KlebanovMQM-\JevickiQN}). 
Considered as a  theory of coupled Liouville and  matter  fields, the 
\QG\
  has much larger symmetry than  each of its two  components. Turning 
on the gravitational field erases all the information about distances and
 instead of correlation functions one obtains `correlation numbers'. 
 One of  the manifestations of this larger symmetry of  \QG \ is the 
so called `ground ring'  integrable 
 structure   discovered  in the early 90's   
\refs{\Witten\WitZw-\KlebPol}.    It has been shown by Witten 
\Witten\ that
 for $c = 1$ matter coupled to Liouville theory,  there exists a ring 
of ghost-number zero  operators, with respect to which the local observables, represented by 
 ghost-number one  vertex operators, form a module. 
 Kutasov, Martinec and Seiberg  \KMS\  have pointed out that the 
 action of  ground ring     leads to to a set of  recurrence 
equations for the   correlation functions  of  the primary fields (the closed string 
`tachyons').
    The ground ring structure  is  particularly interesting also  by the
        fact  \refs{\Witten, \KMS} that it  resembles  the integrable 
structures  observed in the 
        matrix models of \QG \foot{Very recently, new evidences 
        about the deep connection
         between these two integrable structures  were put forward in 
the interesting 
        paper by Seiberg and Shih \SeibergS.}.

   Bershadsky and Kutasov \bershkut\   extended the  ground ring 
    structure  to the case of  \QG\  with boundary, or $c= 1$ open 
 string theory.   In addition to the two bulk ground ring 
generators,  it involves two 
  boundary operators, which generate the `boundary ground ring'.
  In  \bershkut,   the  action of the boundary  ground ring   
   is used to find recurrence relations for the correlation functions
   of boundary  operators, or  open string  `tachyons'.
   The compatibility of these  (over determined)
   relations allows to evaluate all structure constants and all 
correlation 
   functions,  and to reproduce elegantly the results obtained 
previously 
    by Coulomb gas  integrals   
\refs{\Gouli\VDotsenko\DiK\berkut-\TY}.

 Originally  the   ring relations    were   applied  only for the 
 so called   `resonant'
amplitudes,  which are the only non-vanishing ones in a theory with 
vanishing  bulk and boundary  cosmological constants.  
In the full theory, the resonant amplitudes give the residues of the 
`on-mass-shell' poles of the exact amplitudes. 
 It was    conjectured   \Witten\  that  the  perturbations like the 
cosmological term 
   are described by  certain  deformations of the  ground ring 
structure,  
   but    the exact field-theoretical meaning of this conjecture 
   remained obscure.
Only after the   impressive developments in Liouville CFT  from
     the last several years  \refs{\DO\ZZtp\FZZb\PTtwo-\hosomichi}
     it  has been realized  that the  ground ring structure  is 
   an {\it exact}  symmetry of the full CFT
constants 
    \newhat .
     Indeed, the    conformal bootstrap used to derive the exact 
expressions  for  the 
    bulk and boundary  structure constants  in 
\refs{\FZZb\PTtwo-\hosomichi}  
   is  essentially the same procedure as the one used to establish 
the 
   ground ring relations.  The only  technical difference is that 
while the  identities obtained in
     pure Liouville theory follow from the truncated OPE with 
degenerate Liouville fields,    in  \QG\   
      one considers products of degenerate Liouville and matter 
fields.

   In this paper we  consider the boundary ground ring structure in
    Liouville quantum gravity 
   with non-vanishing boundary cosmological constant, 
     or  equivalently, in a 2D string theory in presence of a  FZZT 
brane.
    Extending     the   the technique   developed by Bershadsky and 
    Kutasov \bershkut, we  obtain a set of functional recurrence 
equations for the
    $n$-point boundary correlation functions.
Crucial role in our analysis will play  an  observation made by V. 
Fateev, 
A. Zamolodchikov and Al. Zamolodchikov \FZZb\  about  the  nature   
of the degenerate boundary Liouville fields, which we sketch  
below\foot{
The general validity of this suggestion, which was  proven in \FZZb\ 
only for the two-point function,  follows   from the  formula for the 
three point                                                          
function derived  subsequently in \PTtwo. 
}.

  In  boundary  Liouville  CFT  there are two cosmological constants
 $\mu$ and $\mu_B$,  the bulk and the boundary one.
$\mu_B/\sqrt{\mu}$.   
 The observables   are meromorphic functions 
of $\mu_B$,    with  a branch point singularity  at $\mu_B= - 
\mu_B^0$,
 where    $\mu_B^0\sim\sqrt{\mu}$  is minus  the boundary entropy 
produced by 
  the fluctuations of the Liouville field in the bulk.
  The branch point can be resolved by 
 introducing a uniformization parameter  $s$
 \eqn\deftau{
 \mu_B = \mu_B^0 \cosh (\pi bs)}
 where $b$ is the Liouville coupling constant.
FZZ observed  \FZZb\ 
 that the correctly defined lowest degenerate boundary operators 
 introduce shifts of the  boundary parameter $s\to s\pm ib$ or $s\to 
s\pm  i b^{-1}$
  at the points of the boundary where  they are inserted.
 Therefore, in order to close the conformal bootstrap, 
 one is   led to associate   with a boundary operator two independent 
boundary parameters, 
 $s_{\rm left}$ and  $s_{\rm right}$, labeling  the Liouville 
boundary conditions on  both sides. 
 FZZ derived a  simple functional  equation for the  boundary 
two-point function (see the 
 concluding section of \FZZb), which involves a shift of  one of the 
 two boundary parameters as well as the Liouville charge
 of the two boundary operators.
Liouville fields. 
The boundary two-point function can be obtained as the unique 
solution of this 
equation that satisfies the reflection and duality symmetries.

The functional equation of FZZ can be written as a difference equation
with respect to one of the boundary parameters. In   \KPS\  it was 
found that 
all boundary Liouville structure constants satisfy  
 similar  difference equations.    Interestingly,  these    
equations    have a 
 natural interpretation in   the discrete formulation of \QG\ in 
terms of a 
 gas of loops and lines on a randomly triangulated disc.  
  In the discrete  approach the difference equations are satisfied by 
the 
   $n$-point boundary correlation functions with any $n$, while in  
pure 
   Liouville theory this is so only for $n\le 3$.  Of course, this is 
so because starting with $4$ operators, the 
   contributions of the Liouville and matter  sectors do not 
factorize.

  Our  original motivation for this work   was  
  to give a continuous derivation of the  difference equations for 
the boundary correlators obtained in \refs{\Ibliou, \KPS } within  
the discrete approach.   
  This  naturally led us  to consider  a  deformation of the  
boundary ground ring structure by both bulk and boundary cosmological 
terms.
  Using the  results of    \refs{\FZZb, \KPS} and 
    extending   the   technique developed in \bershkut,
    we  derived a set of  functional 
  recurrence equations for the $n$-point  boundary correlation 
functions
  that   generalize   the recurrence equations  of  Bershadsky and 
Kutasov
 \bershkut \ on one hand,   and the  difference equations  
 obtained in \FZZb\ and \refs{\Ibliou, \KPS }  on the other  hand.  
    These equations completely determine 
 all boundary correlation functions in \QG.

   In this paper we consider the  
realization of 2D string theory on a FZZT brane 
in which the matter field is  a gaussian field  with  pure Neumann or 
Dirichlet boundary conditions. 
 The target space of the gaussian field  represents  the Euclidean
   time direction $x$. The generators of the boundary ground ring, 
being essentially products of Liouville and matter boundary 
degenerate fields,  introduce shifts both in the Liouville boundary 
parameter  $s$ and in the time direction $x$. The time shift  is 
equal to the critical distance  $\Delta x=1$ 
 associated with the Kosterlitz-Thouless point.
In general the  ring relations yield   functional  equations 
which involve shifts  both in the  Liouville and matter boundary 
conditions, 
but in  the  simplest case    we are considering 
the correlation functions  depend  only on the Liouville boundary 
parameter $s$.

The minimal  configuration of FZZT branes 
 that closes  under the action of the boundary ground ring
 is given by an array of 
 FZZT branes   spaced at the critical distance $\Delta x=1$ in the 
time direction.
 We  propose  a matrix model   description for such a background  
based on  a variant of the $\hat A_\infty$ matrix model. In this 
matrix model the boundary ground ring has explicit realization as  
the  ring of the polynomials of 
 of two complex matrix fields.

 \newsec{Some preliminaries}
 
 \subsec{The boundary CFT for the FZZT brane   }
 
\noindent
 We  consider the non-unitary   realization of  Euclidean 2D string 
theory
  by  a Liouville  field $\phi$   and  a gaussian  matter field $\chi$
   with background charge $e_0$.   
  The  background charge is normalized so that the conformal anomaly  
of the matter   field  is  
  \eqn\centc{
 c= 1- 6 e_0^2.
 }
The  disc partition function   of string theory  with  FZZT-type  
 boundary condition \refs{\FZZb, \Teschner} is  defined by by the 
following effective action, 
 which comprises a bulk and a boundary term:
\eqn\actg{\eqalign{
\CA [\chi,\phi]& = \int \limits_\CM
\( {1\over 4\pi}
 [ (\nabla \phi)^2 + (\nabla \chi)^2+ 
 (Q \phi -i e_0\chi ) \hat R 
 ]
+\mu 
e^{ 2b\phi }\)
 \cr
& +
\int\limits_{\p\CM}   \( {1\over 2\pi} ( Q\phi-i e_0\chi )\hat K
+ \mu_B \ e^{b\phi}\) + {\rm ghosts}}} 
where the couplings    $\mu $ and $\mu_B$ are  are  
 referred to as bulk and the boundary
cosmological constants.
The background charges are expressed in
terms of the Liouville coupling constant $b$ as %
 \eqn\bgcharges{ Q= {1\over b}+b, \qquad e_0= {1\over b}-b .  } 
 With the choice \bgcharges\ the two background charges satisfy $Q^2 -
 e_0^2=4$, which is equivalent to the balance of the central charge $
 c_{\rm tot}\equiv c_\phi +c_\chi+ c_{\rm ghosts}=(1+6 Q^2)+(1-
 6e_0^2) -26 =0.  $ 
 We will consider   the generic  situation when  $b$ is not a 
rational number.
 
 After mapping the disc $\CM$ to the upper half
 plane $\{ \Im x\ge 0\}$, the curvature term disappears
\eqn\actUHPQG{ \CA[\phi, \chi] =\int \limits_{\Im x\ge 0}d^2 x\(
{1\over 4\pi} [ (\nabla \phi)^2 + (\nabla \chi)^2 ] +\mu e^{ 2b\phi 
}\)
+\int\limits_{-\infty}^\infty dx \ \mu_B\ e^{b\phi} + {\rm ghosts} }
and the background charges are introduced through the asymptotics of
the fields at spatial infinity \eqn\asyms{
\phi(x, \bar x ) \sim - Q \log |x|^2
,\quad \chi (x, \bar x) \sim - e_0\log |x|^2.
}
The  boundary term encodes a inhomogeneous 
Neumann boundary condition for the Liouville field
\eqn\bcphi{ i(\p-\bar \p) \phi(x, \bar x)  =4\pi \mu_B \ e^{b\phi(x, 
\bar x )}\qquad
(\bar x =x),}
which describes the FZZT brane characterized by the the parameter
 $\mu_B$.  The matter field 
   is assumed to satisfy pure Neumann
 boundary condition:
 \eqn\bcchi{ i(\p-\bar \p) \chi(x, \bar x)  =0 \qquad
(\bar x =x),}
  which is of course  Dirichlet boundary condition for the 
 T-dual field $\tilde \chi (x, \bar x)
  .$
 In this paper we  consider  the simplest  situation where no matter 
screening charges are added.

\subsec{Bulk and boundary vertex operators 
(closed and open string tachyons)}

 \noindent
  The bulk primary fields $\CV_P^{(\pm)} (x,\bar x)$, or 
   left/right moving  on-mass-shell
   closed  string tachyons, are defined as
  the bulk vertex operators
\eqn\newnor{\eqalign{
 \CV_P^{(+)}&= {  1\over \pi}
 \g( bP)\ 
  e^{i(e_0-P)\chi  +(Q- P)\phi}\cr 
 \CV_P^{(-)}&= {  1\over \pi}
 \g\(- \frac{1 }{b}P\) \ 
  e^{i(e_0-P)\chi  +(Q+ P)\phi}
 }
 }
 where $\g(x)\equiv  \G(x)/\G(1-x)$. 
 This normalization removes the external  `leg pole' factors   
in the correlation functions and  is the one to be used  when  
comparing with the microscopic realizations of \QG.   

We will be mainly interested in the correlation functions of the 
boundary fields that can be inserted along the boundary of the world 
sheet.
The boundary primary fields $\CB _P ^{(\pm)} (x)$,  or 
   left/right moving  on-mass-shell open   string tachyons, 
   are defined as the 
boundary vertex operators
 \eqn\normB{\eqalign{
 \CB  _P^{(+)}
   &={1\over \pi} \  \G(2b P)\ 
   e^{ i \(\hf e_0-P\) \chi+\(\hf Q - P\) \phi}\cr
\CB  _P^{(-)}
  & ={1\over \pi}  \ \G\(-\frac{2}{b}P\)\ 
   e^{ i \(\hf e_0-P\) \chi+\(\hf Q + P\) \phi}.
    }
    }
 As any    CFT boundary operator,  the open string tachyon is  
unambiguously  
 defined only after both  left and right boundary conditions are 
specified, which 
 should be done both for matter and Liouville components.
 
 In  a theory  in which the ends of the open strings  freely 
propagate in the 
 two-dimensional space-time ($\mu_B=0$),   both fields satisfy pure 
Neumann boundary conditions.    
In a theory with  FZZT branes,   
the Liouville left and right boundary conditions are determined by 
the 
values $s_1=s^{\rm left}$ and $s_2=s^{\rm right}$ of the 
uniformization parameter
 defined in \deftau . 
 Slightly modifying the  notations of  FZZ \FZZb,  
 will denote such an operator as 
 $^{s_{1}} [\CB  _P^{(\pm)}]^{s_{2}}$. 
   Geometrically the boundary fields   create open string states  
  whose   ends propagate  in two different FZZT branes
  labeled by $s_1$ and $s_2$.

      The   physical observables   can be represented in
two pictures: 
either as (1,1)-forms   and 
  integrated over the the
world-sheet (for the bulk fields) and  1-forms integrated over the 
boundary (for the boundary fields), or 
as BRST-closed 0-forms:
\eqn\twoforms{\eqalign{
\int d^2 x \ \CV_P^{(\pm)} (x,\bar x)\qquad &\leftrightarrow \qquad
{\rm \bf c \bar c}\   \CV_P^{(\pm)} (x,\bar x)\cr
\int d x \ \CB  ^{(\pm)} (x)\qquad &\leftrightarrow  \qquad {\rm  c} 
\ \CB  ^{(\pm)} (x).
}
}
 Here and in the following ${\rm\bf  b, c}$ and ${\rm b, c}$ are  the 
reparametrization
  ghosts respectively in the bulk and on the boundary. 
The second representation is more  appropriate
 as far as the ground ring structure is concerned.

 \subsec{Reflection property}
 
 \noindent
   The tachyons of opposite chiralities are related by the Liouville  
bulk and boundary reflection amplitudes (see Appendix A)
\eqn\reflBul{\eqalign{
\CV_P^{(+ )}&= S_P^{(+-)} \ \CV_P^{(-)}, \cr
^{s _1}  [\CB^{(+)}_P]^{s _2}& =  D_P^{(+-)}  (s _1, s _2) 
\ ^{s _1}  [\CB^{(-)}_P]^{s _2}\cr
}
}

 \subsec{Physical states in \QG}

  For each value of the momentum there is only one 
  vertex operator that corresponds to a physical state. 
  The physical operators are
  \eqn\KPZcl{
  \CV_P=\cases{\CV  _P^{(+)} \qquad  (P>0)\cr
  \CV  _{P}^{(-)}\qquad   (P<0)},\qquad\qquad
  \CB_P=\cases{\CB  _P^{(+)} \qquad  (P>0)\cr
  \CB  _{P}^{(-)}\qquad   (P<0).}
  }
  The  ``wrongly dressed" operators are related to the physical ones 
by the 
  Liouville reflection amplitude.

  \subsec{Degenerate fields}
  
  \noindent
   By degenerate fields in Liouville quantum gravity we will 
understand 
   the gravitationally dressed degenerate matter fields.
    In our case these are the on-shell vertex operators  \KPZcl\ 
    with  degenerate matter components
\eqn\degenPrs{\eqalign{
 \CV_{rs}=    \CV_{r/b-sb}  , \qquad \CB_{rs} = \CB_{\hf(r/b-sb)}
 \qquad (r,s\in\IN).
 }}
The fusion rules for these  fields  are  the same as
 the fusion rules for the degenerate fields  in the matter CFT \BPZ.
 Note that  the  Liouville components of these fields   are not
   degenerate  Liouville fields.

  The operators that  span the ground ring are 
  degenerate with respect to the {\it full}  conformal algebra.  
  Such an operator  is  constructed by  applying  a
   rising Virasoro operator of level $rs-1$
  to a product of matter and Liouville degenerate fields with Kac 
labels $r,s$.
  The   conformal weight of the product is zero:
  \eqn\Deltars{\Delta= 
  {(r/b-sb)^2 - e_0^2\over 4}  +{Q-(r/b+sb)^2 \over 4} 
  +rs-1 = 0.
}

\subsec{Normalization of the bulk and boundary cosmological constants}

\noindent
It is also convenient to redefine the   
cosmological constants    $\mu$ and $\mu_B$   according to
the normalizations   \newnor  \ and \normB\ of the vertex operators.
The new bulk and boundary  cosmological constants  $\Lambda$ 
and $z$  are defined as the couplings of the operators  
$\CV^{(+)}_{e_0}$
and $\CB  ^{(+)}_{e_0/2}$ respectively:
\eqn\intBB{
\mu    e^{2b\phi} =
\Lambda\  \CV^{(+)}_{e_0},
\quad \mu_B \ e^{b\phi}= z\  \CB  ^{(+)}_{e_0/2}.  
}
 This gives
 \eqn\Mmu{
{\Lambda\over \mu}  =\pi \g(b^2) ,
\qquad {z\over \mu_B} =  {\pi\over \G(1-b^2)}.
} 
The  uniformization map \deftau\   now  reads%
\eqn\zofM{
 z = M\cosh (\pi b s)  , \qquad M= \sqrt{\Lambda}.
 }
Here we introduced the constant $M$ whose physical meaning is that it
is equal to minus the boundary entropy due to the fluctuations of the
Liouville field\foot{When comparing with the discrete approach, the
cosmological constants $\L$ and $z$ are to be identified with the
corresponding quantities in the microscopic formulation of the strings
defined on Dynkin graphs \Idis.  The constant $M$ equals half the loop
tension in the loop gas description of 2D string theory \refs{\Idis,
\KKloop}.  The uniformization parameter $s$ is defined as in \FZZb; it
is related to the parameter $\t$ of \Idis\ by $\t = \pi b s$.}.

\subsec{The self-duality property of Liouville quantum gravity}

\noindent
All correlation function in Liouville theory are invariant 
w.r.t.  the  substitution \FZZb\ 
$$ b\to \tilde b = 1/b, \qquad \L\to \tilde\L, \qquad M\to \tilde M$$
where 
 \eqn\DUALTY{ \tilde \Lambda =
 \Lambda^{1/b^2} ,  \qquad \tilde M= M^{1/b^2}.
} 
The boundary parameter $s$ is self-dual.  The duality transformation
for the correlation functions follows from the one 
of left and right chiral tachyons \eqn\TDU{ \tilde \CV ^{(\pm)}_{P} =
\overline{ \CV ^{(\mp)}_{-P}} , \qquad ^{s_1}[\tilde \CB ^{(\pm)}_{P}
]^{s_2} =\ ^{s_1}\![ \overline{ \CB ^{(\mp)}_{-P}}]^{s_2}, }
where the bar means   complex conjugation.

 It is clear from \TDU\ that the non-unitary CFT under consideration
 is not self-dual in the strict sense.  The duality transformation
 interchanges the states of a pair of CFT characterized by matter
 background charges $e_0= \frac{1}{b}-b$ and $\tilde
 e_0=b-\frac{1}{b}=-e_0$.  The duality becomes a symmetry (a chirality
 flip) only when it is accompanied by a complex conjugation. 
 

 \newsec{The  (bulk)    ground ring }

 \noindent
 We start with a  short review of  the   ground ring structure
for a theory defined on  the sphere ($\p\CM=0$). 
 The ground ring operators are obtained by applying raising operators
of level $rs-1$ to the product of  two degenerate matter and 
Liouville fields
with Kac labels $r,s$.  The resulting operators have conformal 
weights $\Delta=\bar\Delta=0$.
The    ring is generated by the lowest o  two perators \Witten
\eqn\bulkGR{
\eqalign{
a_+=&-|{\rm\bf bc}- b\p_x(\phi-i \chi)|^2
e^{- b^{-1}(\phi+i\chi)}\cr
 a_-=&-  |{\rm\bf bc}- b^{-1} \p_z(\phi+i \chi)|^2
 e^{ -b(\phi -i\chi).}
 }
 }
where  ${\rm\bf  b, c}$  are the
reparametrization ghost and anti-ghost fields.
The  ground ring is spanned on the polynomials
 $(a_+)^m(a_-)^n$ with  $m,n\in \IZ_+$.
 In the case of non-rational $b^2$ the ground ring contains no other 
relations
 and has an infinite number of elements labeled by the integers 
$r,s\ge 1$.    

 A crucial property of the operators $a_\pm$ is that their
 derivatives $\p_x a_\pm$ and  $\p_{\bar x} a^\pm$ are
    BRST exact: $\p_x a_- = \{Q_{\rm BRST}, {\rm\bf b}_{-1} a_-\}$. 
    Therefore, any amplitude that
involves $ a_\pm$  and other BRST invariant 
operators does not depend on the
position of $ a_\pm$.
This property allows to  write recurrence equations for the 
correlation functions from the  OPE   of $a_\pm$ and the other 
BRST-invariant fields.

\subsec{The limit of  of free fields  ($\L=0$)}
\noindent
The vertex operators   \newnor\  form a module under the
  ground ring:
\eqn\actaa{
 a_{+} \CV_P^{(+)} = -
   \CV_{P+  {1\over b } }^{(+)} , \qquad
   a_{-}  \CV_P^{(-)} =- 
   \CV_{P-  {b} }^{(-)} .
}
  and also 
  \eqn\actaab{
  a_{+} \CV_P^{(-)} = a_{-} \CV_P^{(+)} =0.
  }
  Both relations  {\actaa}\ and  {\actaab}  follow from the free 
field OPE\foot{The normalization of the fields is determined by the 
   action \actUHPQG\ and is the same as in \FZZb.} and 
  are true up to commutators  with the BRST charge.
 While the first one survives in the  correlation functions,
 the second one receives non-linear corrections. 
 Due to the contact terms, the last relation  should be 
  modified   in presence of an integrated  vertex  operator 
  to  
  \eqn\aplus{
    a_+  \CV^{(-)}_P\int d^2 z\   \CV^{(-)}_{P_1}
=  
\CV^{(-)}_{P+P_1+ b}
}
\eqn\aminus{
 a_-  \CV^{(+)}_P\int d^2 z \   \CV^{(+)}_{P_1}
=   
\CV^{(+)}_{P+P_1-b\inv.}
}
These relations  imply
 a set of recurrence equations  \refs{\KMS, \kachru, \bershkut}, which
   determine completely the resonant   
tachyon amplitudes (see Appendix B).

\subsec{$\L\ne 0$}

 \noindent 
The first   relation \actaa\  survives the perturbation with  $\L\ne 
0$,
while the second  relation  {\actaab}  gets  deformed.
  The deformation   is linear  
  in $\L$ and can be calculated pertubatively
  by considering  a first order insertion of the 
Liouville interaction 
$
-\mu  \int e^{2b\phi} =
- \L \int\CV^{(+)}_{e_0}$
and then 
applying  {\aminus} :
   \eqn\actaba{
a_-  \CV^{(+)}_P 
=   - \L\ 
\CV^{(+)}_{P-  b }.
}
This procedure has been justified by various  self-consistency checks 
in  Liouville theory \refs{\DO, \ZZtp}.
 By the same argument one finds the action of  the operator $a_+$.
  This time one should consider a linear insertion of the 
    dual Liouville interaction 
$  -\tilde \mu  \int e^{2\phi/b} =
- \tilde \L \int \CV^{(-)}_{-e_0}$. The result is
 \eqn\actabd{ 
a_+  \CV^{(-)}_P =    - \tilde \L\ 
\CV^{(-)}_{P+1/  b }, \qquad \tilde \L =  \L^{1/b^2}. 
}
 The deformed ring relations allow to derive an (over determined) set 
of   recurrence equations for the bulk correlation functions (see 
Appendix B).

 Here we consider the simplest situation where the matter field obeys 
 $U(1)$ fusion rules.    In  general   the    
   free-field representation  of \QG\   should be  completed 
  by   adding   to the effective  action 
\actg\ 
  the two matter screening  charges  
  $$Q_+= -\frac{1}{ \g (-1/b^2)}\int e^{2i \chi/b}  \quad {\rm  and }\quad 
    Q_- =  -\frac{1}{  \g(-b^2)} \int e^{-2ib\chi}.$$
    \noindent
    Then eqns. \actaba\ and \actabd\ will further deform to
     \eqn\acSC{
a_-  \CV^{(+)}_P 
=   - \L\ 
\CV^{(+)}_{P-  b } +\CV^{(+)}_{P+  b } 
}
    \eqn\actSC{ 
a_+  \CV^{(-)}_P =    - \tilde \L\ 
\CV^{(-)}_{P+1/  b } +  \CV^{(-)}_{P-1/  b }.
}
 The  ground ring structure  in presence of screening charges has 
been considered
 in    \refs{
 \GovJcf, 
 \GovLast}.

 \subsec{The case $b=1$}

\noindent
 The case $b=1$ plays a special role \Witten. 
 Consider the action of the product $a_+a_-$:
 $$  \eqalign{
 a_+a_- \CV^{(+)}_{P}& = \L \ \CV^{(+)}_{P+e_0},\cr
 a_+a_- \CV^{(-)}_{P} &= \tilde\L \ \CV^{(-)}_{P+e_0}.
 }
 $$
 When $b=1$ we get,  since  $\L=\tilde \L$ and $e_0=0$,  
 \eqn\eqFS{
  a_+a_- = \L\qquad (b=1).
  }
  This  equation has a direct interpretation in the  profile of the 
Fermi surface 
  in the corresponding  large $N$ 
  matrix model  \Witten.
  This analogy can be pursued further to a generic perturbation by 
  bulk tachyons and relate the ground ring structure and the Toda 
integrable structure in the 
  matrix realization of the $c=1$ string theory 
\refs{\DMP\MukhiImbimbo\KKK\AKK\Ise-\Iflows}.

\newsec{The boundary   ground ring}

\subsec{The boundary ground ring for  pure Neumann boundary 
conditions ($\mu_B=0)$}

\noindent
 We  first introduce the   boundary ground ring  for 2D open strings 
with  
 of pure Neumann boundary conditions for both coordinate fields, 
$\phi$ and $\chi$,
 following  Bershadsky and Kutasov \bershkut.
 The  two generators are defined, following the same logic as in the 
bulk case, as
 \eqn\APAM{ \eqalign{
A_+
& 
=- [{\rm bc} - \hf b\p_x(\phi -i\chi) ] \ e^{-\hf b\inv  
(\phi+i\chi)}\cr
A_- 
&  
=- [{\rm bc} - \hf b^{-1} \p_x(\phi +i\chi) ] \ e^{-\hf b 
(\phi-i\chi)}.
}
}
The two operators are related by a duality transformation combined 
with complex conjugation:
\eqn\conjA{
\tilde A_+ = \overline{A}_-, \qquad
\tilde A_- = \overline{A}_+.
}
They are BRST closed: $ \p_x A_\pm = \{ Q_{\rm BRST} , {\rm b}_{-1} 
A_\pm\}$ 
and have $\Delta=0$. The  open string tachyons \normB\  
form a module with respect to the 
ring generated by these two operators. The
action of the   ring on the tachyon modules is 
generated by the relations
 \eqn\ACTAA{
 A_{+} \CB  _P^{(+)} = 
  \CB  _{P+  {1\over 2 b } }^{(+)} ,\qquad 
   A_{-} \CB  _P^{(-)} = 
  \CB  _{P-  {b\over 2} }^{(-)} 
  }
  and
  \eqn\ACTAAB{
  A_{+} \CB_P^{(-)} = A_{-} \CB_P^{(+)} =0  .
  }
   If one exchanges the places of $A$ and $V$, there will be minus on 
the rhs, since $A$ is odd and $V$ is even wrt  $x\to -x$:
 \eqn\revrsd{
 \CB  _P^{(+)} A_{+}  =-
  \CB  _{P+  {1\over 2 b } }^{(+)} ,\qquad 
\CB  _P^{(-)}   A_{-} = -
  \CB  _{P-  {b\over 2} }^{(-)} .
 }
   Again, the first relation \ACTAA\ is exact and the second relation
   \ACTAAB\  gets deformed in presence 
   of integrated boundary tachyon fields  \bershkut :
   \eqn\nonAr{ A_- \CB  _{P_1}^{(+)} \int dx \CB  _{P_2}^{(+)}=
   {1\over \sin 2\pi b P_1 }\   \CB  _{P_1+P_2-  {1\over 2b}}^{(+)}.
      }
   Since on the boundary the ordering is important, there is a second 
relation
    \eqn\nonAl{
    \int dx \CB  _{P_1}^{(+)} A_- \CB  _{P_2}^{(+)} =
    {\sin 2\pi b (P_1 +P_2)\over \sin 2\pi b P_1\ 
    \sin 2\pi bP_2}\ 
     \CB  _{P_2+P_1-  {1\over 2b}}^{(+)}.
   }
   The coefficients in \nonAr\ and \nonAl\ are obtained 
   as standard Coulomb integrals
   (see for example \TY).
  Similarly,  one finds  for $A_+$
   \eqn\nonApr{A_+
   \CB  _{P_1}^{(-)} \int dx\CB  _{P_2}^{(-)}=
   {1\over \sin {2\pi\over b}   P_1 }\  
    \CB  _{P_1+P_2+  {b\over 2}}^{(-)}
   }
  and
    \eqn\nonApl{
    \int dx \CB  _{P_1}^{(-)} A_+\CB  _{P_2}^{(-)} =
    {\sin {2\pi\over b}  (P_1 +P_2 )\over 
    \sin {2\pi\over b} P_2 \ \sin {2\pi\over b}  P_1 }\ 
     \CB  _{P_1+P_2+ {b\over 2}}^{(-)}.
        }
        Using these relations,  Bershadsky and Kutasov
        obtained  recurrence equations for 
        the  $n$-point  open string amplitudes (Appendix C).

\subsec{The boundary ground ring  in presence of  FZZT branes}

\noindent
The ring relations can be generalized to the case 
of  nonzero  boundary  cosmological constant 
 using the results of  the  paper by V. Fateev, A. Zamolodchikov 
 and Al. Zamolodchikov \FZZb.   
 The crucial observation made  by the authors of \FZZb\  and  later 
confirmed by 
 B. Ponsot and  J. Teschner \PTtwo\  is that a 
  level-$n$ degenerate boundary Liouville field  
  $e^{-nb\phi}$  has vanishing null-vector and therefore a truncated 
OPE with the other primary fields if either $s_{\rm left} - s_{\rm 
right } = i b k$ or 
  $s_{\rm left} + s_{\rm right } = i b k$, with $k = -n/2, 
-n/2+1,..., n/2$.
  By the duality property  of Liouville theory, the degenerate 
boundary  fields 
  $e^{-n \phi/b}$ exhibit a similar property    
  with $b$ replaced by $1/b$.   No direct proof is supplied for 
  this statement, but it was shown to be consistent with the exact   
results obtained in boundary Liouville theory. 
 In particular,    the above property leads to a pair of
   functional equations for the Liouville  boundary reflection 
amplitude, whose
unique solution coincides with the result of the standard conformal 
bootstrap.

  According to \FZZb, 
 the  operators \APAM\  should be   defined  by\foot{We omit the 
 boundary conditions for the matter field.
 In  our   case of  gaussian  field  $\chi$ with pure Neumann 
boundary conditions,
 those are given by the values of the dual field $\tilde \chi$ on 
both sides.
 The momentum conservation yields $\tilde\chi_{\rm right} - 
\tilde\chi_{\rm left} 
 = \pm b^{\pm1}$. In calculation expectation values, the final result 
will 
 not depend on the  only residual  parameter, the global mode of 
$\tilde\chi$.
 However, if screening charges are allowed,  the matter boundary 
conditions  matter.}
 \eqn\APMnew{\eqalign{
 A_+\quad &\to\quad  ^s[A_+]^{s \pm i /b}\cr
 A_-\quad &\to\quad  ^s[A_-]^{s \pm i b}
 }}
and the relations  \ACTAA,  \nonAr,  \nonApr\    and  \nonApl\   
understood as
    \eqn\leftA{
^{s_1}[ \CB  _P^{(-)} ] ^{s }  [A_-]^{s \pm i  b} = -\ 
 ^{s _1} [\CB  _{P-  {b\over 2} }^{(-)}]^{s \pm ib} }
  etc.
  As in the bulk case, the  relation \ACTAAB\ is deformed  by a 
linear  insertion of 
 the boundary Liouville  interaction.  However, in this case the situation is 
more subtle 
since the boundary interaction depends on the point it is inserted.
  We  use the fusion rules  that follow from {\nonAr} and  {\nonAl} 
  where one of the operators is 
  the Liouville boundary interaction
$-z(s ) \times \int\CB  _{e_0/2}^{(+)}$.
     The result is
    \eqn\funceq{
  ^{s_1}[  A_- ]^{s _{1}\pm i  b}
 [\CB  _{P } ^{(+)}]^{  s _2}
 = C_\pm   \times \  ^{ s _{1} }[\CB  _{P -{b\over 2}} ^{(+)}]^{ s 
_2}}
 with the coefficient  $C^\pm$ given by the sum of the contributions 
of the 
 three possible insertions of  $-z(s )\times \int\CB  _{e_0}^{(+)}$  
with          
  $s = s _1, s _1\pm i b$ and $s _2$  (fig.1):
 \eqn\cccc{
C_\pm  = - M  \( \cosh (\pi b s _1)  {\sin2\pi b(\hf e_0+P )\over
 \sin(\pi b e_0)\sin (2\pi bP)}
 + {\cosh [ \pi b (s _1\pm ib)]\over \sin (\pi b e_0)}
 +{\cosh (\pi b s _2)\over
 \sin(2\pi bP)}\).
 }
 The coefficient  $C_{\pm}$ can be nicely written as  
  \eqn\CPM{\eqalign{
  C_\pm &= -{M} \ {\cosh[ \pi b (s _1\pm 2  i  P) ]+ \cosh(\pi b s _2)
  \over \sin 2\pi b P}\cr
  &=-M{\cosh\[\pi b\({s_1+s_2\over 2 } \pm iP\)\]
  \cosh\[\pi b\({s_1-s_2\over 2 } \pm iP\)\]\over  \hf \sin 2\pi b P}
 .}
}

    \epsfxsize=120pt
   \vskip 20pt
   \hskip 100pt
   \epsfbox{ 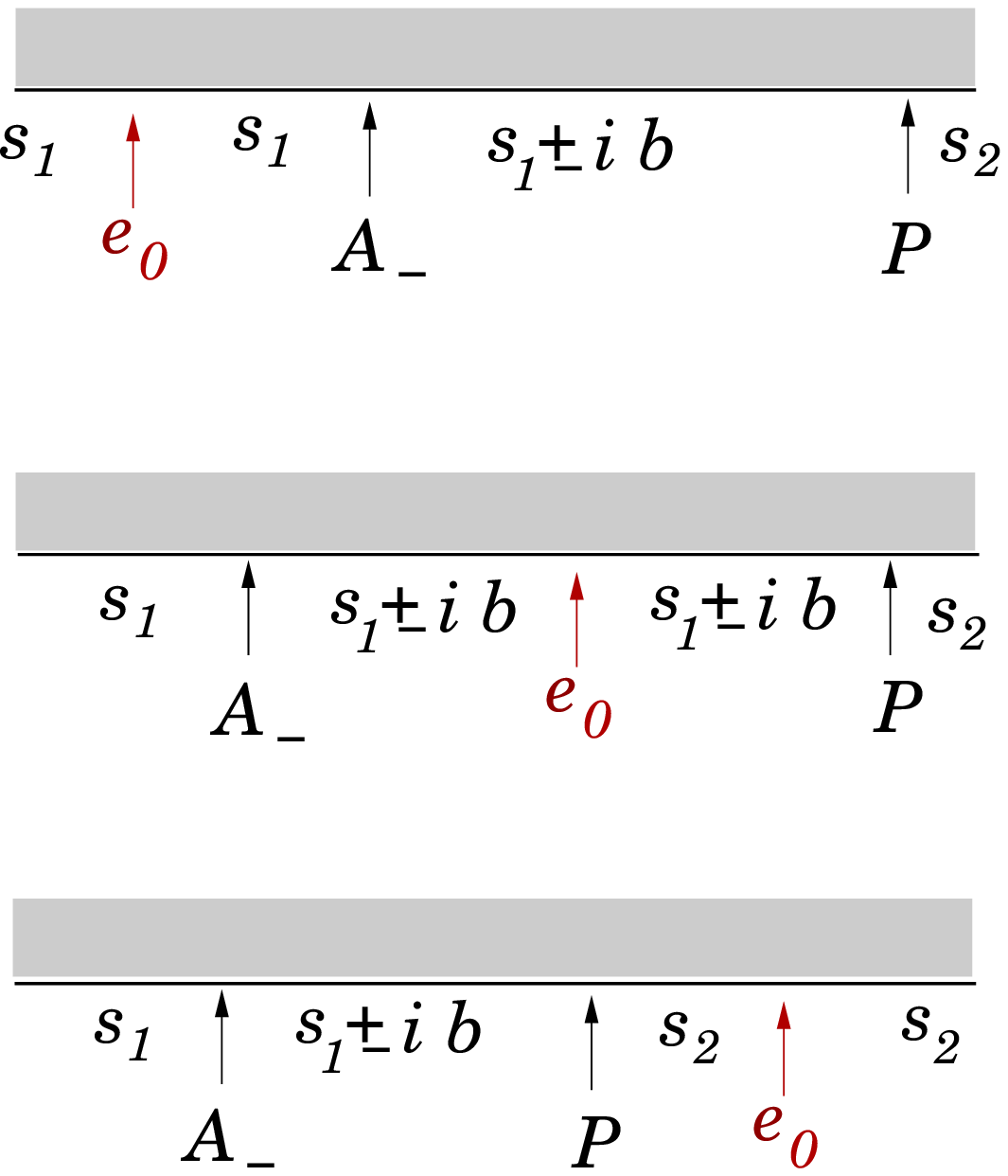   }
   \vskip 5pt
   
   \centerline{\ninepoint 
   Fig. 1:   The three possible insertions of the boundary Liouville  
   interaction.    }
    
    \vskip  20pt

   \noindent
 Similarly one  finds for the action of  $A_+$  on $B^{(-)}_P$ 
     \eqn\funceqd{
  ^{s_1}[  A_+]^{s _{1}\pm i /b }
 [\CB  _{P } ^{(-)}]^{  s _2}
 = \tilde  C_\pm\ \times\  ^{ s _{1} } [\CB  _{P +{1\over 2b}} 
^{(+)}]^{ s _2}}
 with 
  \eqn\cct{\eqalign{
  \tilde C_\pm &=-M^{1/b^2} \
   {\cosh[\pi   \( {s _1} \pm 2i P\) /b ]+ \cosh(\pi {s _2/ b})
   \over \sin 2{\pi\over b}  P}\cr
  &=-M^{1/b^2} \
  { 
  \cosh\[ {\pi\over b} \( {s _1+s _2 \over 2   } \pm i \pi  P\) \]
     \cosh\[ {\pi\over b} \( {s _1-s _2 \over 2   } \pm i \pi  P\) \]
  \over \hf  \sin (2{\pi\over b}  P)}
  .}
}
Note that the fusion coefficients are symmetric with respect to the 
change 
the sign of any of the boundary parameters. This is a general 
property of all
correlation functions and comes from the fact that    the $s$-space 
is an orbifold:  the points $s$ and $-s$ should be identified.

  \subsec{Functional   equations for the correlation functions
  of boundary primary fields}

   \noindent
    Consider  a  boundary correlation function of the form  (fig.2)
  \eqn\defWP{
 W_{P_1, P_2, ... }(s _1, s , s _2,s_3, ...) 
 =\< ^{s _1}[\CB _  {P_1} ]^{s } [ \CB  _{P_2}  ] 
 ^{ s_2 }[\CB  _{P_3}  ]^{s _3} ...\>}
The realization of the physical boundary fields 
depends on the sign of the momenta and 
is given by \KPZcl. The amplitude \defWP\ is by construction 
symmetric in its arguments. It is   analytic in the boundary 
parameters $s, s_1,...$  but not in the target space momenta. 
By the momentum conservation the correlation function 
is zero  unless $\sum_k (\hf e_0 - P_k) = e_0$.
   Equation 
   \TDU\   implies   the identity
   \eqn\revss{ W_{P_1, P_2, ... }
   =  \overline{W}_{-P_1, -P_2, ... }  
   = W_{-P_1, -P_2, ... }  .
  }

 Let us assume for definiteness that
 $P_1<0$ and $P_2, P_3 >0$. 
 Then the amplitude \defWP\ is
 realized as
 \eqn\defWPa{
 W_{P_1, P_2, P_3,  ... }(s _1, s , s _2,s_3, ...) 
 =\< ^{s _1}[\CB _  {P_1}^{(-)}]^{s } [ \CB  _{P_2} ^{(+)}] 
 ^{ s_2 }[\CB  _{P_3} ^{(+)}]^{s _3} ...\>}
 By the   symmetry  \revss\
  the amplitude 
   can be 
 written as 
 \eqn\defWPd{
  \overline{W}_{P_1, P_2, P_3, ... }(s _1, s , s _2,s_3, ...) 
 =\< ... ^{s _3}[\CB _  {-P_3}^{(-)}]^{s_2 } [ \CB  _{-P_2} ^{(-)}] 
 ^{ s}[\CB  _{-P_1} ^{(+)}]^{s _1} \> .
 }
  We will use  these two representations  to derive two independent 
functional identities 
 for  $W$.

  \epsfxsize=100pt
 \vskip 5pt
 \hskip 125pt
 \epsfbox{  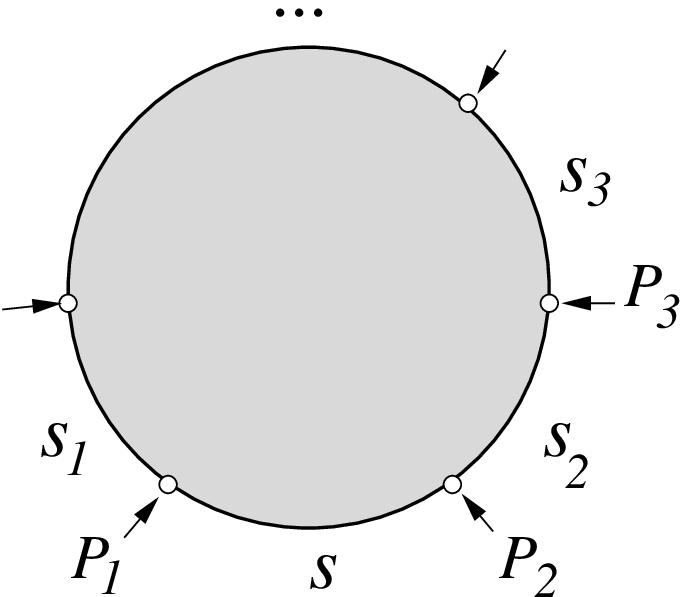  }

 \centerline{\ninepoint 
 Fig. 2:   Our convention for the ordering of the boundary parameters
 and momenta.  }
   \vskip 15pt
  
 \noindent
 Consider the   auxiliary correlation function 
 $F$  with an operator $A_-$ inserted in the r.h.s. of \defWP:
 $$F =
 \< ^{s _1}[\CB  _{P_1+{b\over 2}}^{(-)}]^{s }[  A_- ]^{s \pm ib}
[  \CB  _{P_2}^{(+)} ]^{  s _2}[\CB  _{P_3} ^{(+)}]^{  s _3}
 ...\>.
 $$
 Let us assume that $P_1<-{b\over 2}$ and $P_2> {b\over 2}$. 
 Then, evaluating $F$  by  \leftA\ and by   {\funceq} 
 and equating the results\foot{Our  notations do   
 not distinguish between integrated 
 and non-integrated fields;  
  the rules are the same as in Appendix C.}
   we get the following functional  identities
 for the correlation functions of   three or more  fields:
   \eqn\FUE{\eqalign{
&   \sin\( 2\pi b  P_2\)       W_{P_1
, P_2 ,P_3,  ... }(s _1, s \pm ib  , s _2,s_3, ...)   =\cr 
&2M\cosh\[\pi b\(\frac{s+s_2}{2 } \pm iP_2\)\]
  \cosh\[\pi b\(\frac{s-s_2}{2 } \pm iP_2\)\] 
  W_{P_1+{b\over 2}, P_2-{b\over 2},P_3, ... }(s _1, s   , s _2,s_2, 
...)\cr
& 
+
W_{P_1+{b\over 2}, P_2+P_3 -{1\over 2b}, ... }(s _1, s   ,s_3, ...).
}
}
The last term represents a correlation function with 
one  operator less.   

  A dual equation can be obtained in the same way by evaluating the 
  complex conjugated function $\overline{F}$ using
   the representation \defWPd \ and the relations \conjA:
   \eqn\FUED{\eqalign{
&   \sin\( \frac{2\pi}{b }  P_2\)       W_{P_1
, P_2
,P_3,  ... }(s _1, s \pm \frac{i}{b}  , s _2,s_3, ...)   =\cr &
2M^{1/b^2} \cosh\[\frac{\pi}{b }  \(\frac{s+s_2}{ 2 } \pm iP_2\)\]
  \cosh\[\frac{\pi}{b }  \(\frac{s-s_2}{ 2 } \pm iP_2\)\]
W_{P_1+{1\over 2b}, P_2-{1\over 2b},P_3, ... }(s _1, s   , s _2,s_2, 
...)\cr
& 
+
W_{P_1+{1\over 2b}, P_2+P_3 -{b\over 2}, ... }(s _1, s   ,s_3, ...),
}
}
where we assumed that $P_1<-{1\over 2b}$ and $P_2, P_3 >{1\over 2b}$.
%
These identities    form a set of (over-determined)
 functional recurrence equations that generalize  the recurrence 
equations 
of    \bershkut. 

   Taking the differences in \FUE\ and \FUED\ 
  one finds a pair  of {\it homogeneous}  difference equations which 
do not have analog in the   $\mu_B=0$ limit.  
    We write them  in operator form:
 \eqn\FUim{\eqalign{
 &\[  \sin ( b \p_s)-M \sinh (\pi bs  )\
  e^{{b\over 2}(\p_{P_1} - \p_{P_2})}\]
 W_{P_1, P_2,P_3,  ... }(s _1, s , s _2,s_3, ...)   =0 
}
}
and
 \eqn\FUimD{\eqalign{
 &\[  \sin ( \frac{1}{b }  \p_s)-M^{1/b^2} \sinh (\frac{\pi}{b } s  )\
  e^{{1\over 2b}(\p_{P_1} - \p_{P_2})}\]
 W_{P_1, P_2,P_3,  ... }(s _1, s , s _2,s_3, ...)   =0 .
}
}
where it is assumed that $P_1< - {b\over 2}$ and $P_2>{b\over 2}$.
 The difference equations \FUim\ and \FUimD\ 
 have been first obtained     in the 
microscopic  formulation of \QG\ as a gas of loops and lines on the 
randomly triangulated disc  in   \refs{\Ibliou, \KPS}.

    Functional equations for all values of the momenta  can be 
obtained 
    by the following rule \refs{\Ibliou, \KPS}.
  If  the sign of one or  both  momenta  changes after the shift,
one should  apply the reflection property with the
amplitude $D^{(+-)}_P(s_1,s_2)$ given in Appendix A.  
The details are explained in \KPS.
 
 The functional equations \FUE\ and \FUED\  can be used to evaluate 
the 
 two- and one-point functions as well as the disc partition function
  on the FZZT brane.
  These quantities are difficult to calculate directly  in CFT 
because of the 
 subtleties associated with the residual global conformal symmetry.
 The calculations are presented in  Appendix D.
 

\newsec{Matrix model description}

  \noindent
   The old  idea of Polchinski that  matrix models for non-critical 
strings  
 describe backgrounds populated  by a large number of
  unstable D-branes was recently given a concrete shape  in 
\refs{\mgv\MartinecKA\KlebanovKM\McGreevyEP\SchomerusVV\AlexandrovNN\Sennew\topint-\KMSnew}.

The boundary state of such a D-brane is a product of
the ZZ boundary state \ZZPseudo\ for the Liouville field, localized at
large $\phi$, and a    boundary state for the matter field 
corresponding to some local boundary condition.
For example the standard matrix model for the $c=1$ string, the 
Matrix Quantum Mechanics (MQM),  describes the case of 
pure    Neumann boundary condition for the matter field (or 
equivalently 
pure Dirichlet 
boundary condition for the dual field).

In this section we will show, using the results of 
 \refs{\Ibliou, \KPS},   that there exists a
neat  matrix model   interpretation of   the  
  boundary ground ring relations.
 The   matrix model   dual to the world sheet theory \actg\ 
     is an infinite matrix chain, 
 which can  be viewed as a special discretization of  MQM.
 The logic is the following. First, as it was shown in \refs{\Ibliou, 
\KPS}, 
  the difference equations \FUim\ and \FUimD\   
  have  straightforward  interpretation 
 in   the  loop gas  formulation of  non-critical string theories 
\refs{\ADEold, \Idis, \KKloop}
in which formulation the string path integral is   defined 
microscopically as 
a sum over   planar graphs  embedded in a   Dynkin diagram
of  $ADE$ or $\hat A\hat D\hat E$ type.
  This allows to find  the target-space equivalent of a degenerate 
boundary field.
 Second, the loop gas formulation  can be restated in terms of the 
 $ADE$ and $\hat A\hat D\hat E$  matrix  chains studied in
  \refs{\adem\CMM\Icar\higkos\Iham-\Iopen}.    
   The matrix model description of the  world sheet theory \actg\ 
   is provided by the $\hat A_\infty$ matrix chain.
   
    It is natural to think of the  $\hat A_\infty$  matrix  chain as  a marginal  perturbation of  MQM by a periodic  potential
   $\sim  \cos (2\pi x)$ with the effect that each 
    D0-brane decomposes into an array of  D-instantons 
   associated with  the minima  $x\in \IZ$ of the 
potential\foot{Another possible matrix model description of such a 
perturbation is  proposed in \XID. }.
    The initial state of such a system is  an array of FZZT branes 
    placed at  the sites of the $\hat A_\infty$ Dynkin graph ($N$  
branes per site).

 

 \subsec{Loop gas and  branes}
 
 \noindent
   The continuum limit  of   the  loop gas   
   model can be  given the following interpretation  within the 2D string theory.
     It  describes     the  evolution of    closed strings  in 
presence  
of    an  array of  FZZT branes  spaced  at  the   critical distance $\delta x=1$
in the Euclidean time direction.   
 The      background charge for the matter field  is 
introduced  by
 boosting the   array of D-branes with momentum   $e_0$.   
 The target space of such a string theory   is   a  direct 
product of the Liouville direction and the  discretized time direction
\eqn\chix{\tilde \chi = \pi x b, \quad x\in \IZ,}
in which only the  points 
 of the array  are retained. 
   The string world sheet   is divided into domains  
associated with
   different     branes.   It is  assumed that  only  jumps $x\to x\pm 1$
between two nearest-neighbor   branes are allowed. This  leads to 
the geometrical picture  of a gas of  
      non-intersecting loops, representing   
      the domain walls on the  world sheet (fig.3).

       \epsfxsize=290pt
      \vskip 20pt
      \hskip 10pt
      \epsfbox{ 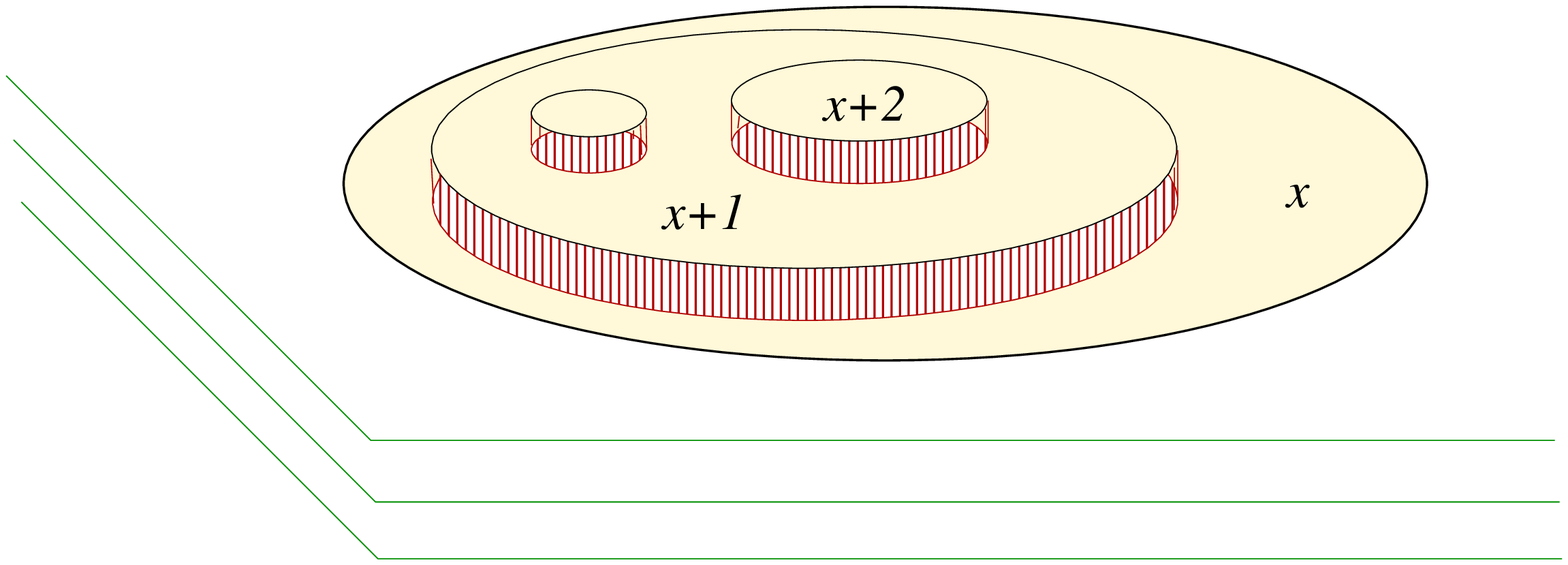   }
      \vskip 5pt
      
      \centerline{\ninepoint Fig.3: A   disc world sheet   with boundary in the 
      $x$-th brane. }
       
       \vskip  10pt

      \noindent
      Creation of a domain wall costs energy  which is proportional 
to the 
      length of the domain wall and which is taken into account by 
the loop tension $M$.   The    statistical weight of each domain 
wall  is $e^{-2M\ell}$  where
     $\ell = \int_{\rm _{^{domain}_{\ \ wall} }}\hskip -15pt 
      e^{b\phi}  $  is the   length of the 
  domain wall.
       The   loop gas    describes a 
     sector of the   world sheet theory \actg\  containing  certain class of 
     degenerate   fields  \degenPrs \ (for a recent discussion see \KPS). 
     Let us remind that the field $\chi$ satisfies Neumann boundary 
     condition and that the dual field $\tilde \chi$ satisfies 
     Dirichlet boundary condition and that the lowest degenerate
       boundary   operator  $\CB_{12}$
      is realized as a {\it magnetic}  operator  with magnetic charge $P=  b/2$.
      The  operator   $\CB_{1,2}$  creates a    jump of the boundary to an adjacent  brane  in the $\tilde \chi$ space.
   Geometrically this  operator   is described by  a   line starting at
   some point on  the boundary. Similarly the 
   degenerate boundary operator  $\CB_{1,n+1}$ with  magnetic charge
   $P=  nb/2$ creates
   a jump by $n$ steps and is described by  $n$   lines  starting at 
   given point on the boundary.  The correlation function of two such 
   operators is a sum of world sheets as the one shown in fig. 4.

    \epsfxsize=250pt
   \vskip 20pt
   \hskip 50pt
   \epsfbox{  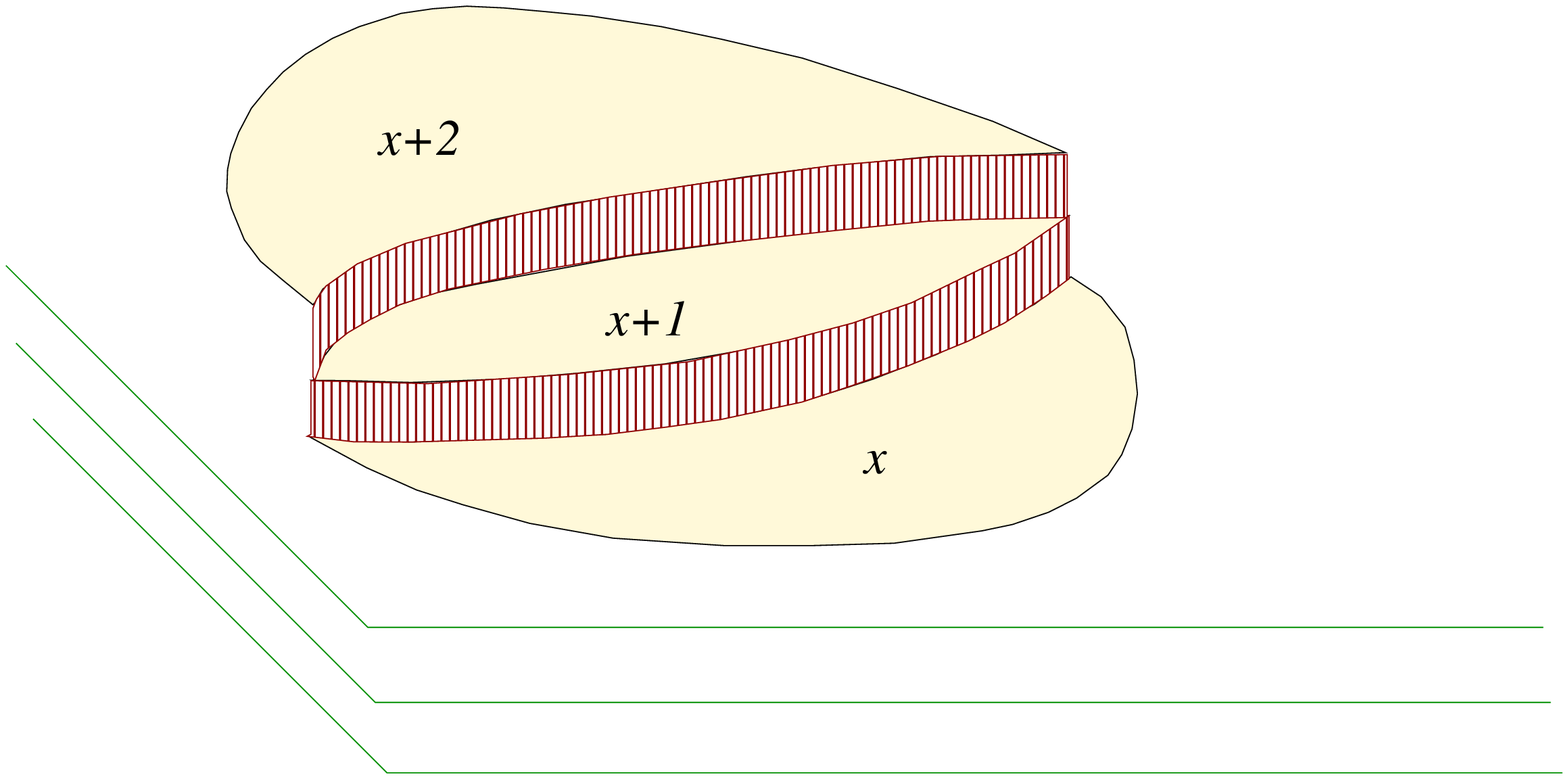  }
   \vskip 5pt
   
   \centerline{\ninepoint 
   Fig.4 :  A world sheet   with two boundary degenerate fields $\CB_{1,2}$.
    }
    
    \vskip  20pt

   \noindent
   
    The  degenerate boundary operators $\CB_{n+1, 1}$ 
    with $n=1,2,...$ can be constructed as 
     {\it electric}  operators interpolating between two Neumann boundary 
    conditions for the field $\chi$. The  Neumann boundary 
    condition is characterized by the global mode
   representing the increment  $\chi = \int d\chi$ along the boundary.
    The action of these operators closes in the discrete set  of boundary 
    conditions characterized by
        \eqn\chixchit{  \chi = \pi x/b, \quad x\in \IZ.}
        
The   {\it duality}  symmetry $b\to 1/b$ exchanges the $\chi$ and $\tilde \chi$ 
 spaces. In the dual   loop gas model   the time coordinate $\chi$ is
  discretized with spacing $\pi /b$.  
   The weight of a domain wall is 
 $e^{-\tilde 2 M\tilde \ell}$,
 where  
 $\tilde \ell = \int_{\rm _{^{domain}_{\ \ wall} }}\hskip -15pt e^{\phi/b} dx$.
 The dual loop tension  $\tilde M = M^{1/b^2}$ is  the dual boundary cosmological constant  for the theory \actg.

 Now let us  see what is the  meaning of  the operators of the boundary ground
  ring in the loop gas model. 
  Geometrically the degenerate boundary operator with 
  momentum $\CB_{n+1,1}$ is obtained from $\CB_{n,1}$
  by adding an extra line. This reminds the fusion rules for the 
  matter fields. However, the process of adding a line cannot be interpreted
  directly  as a fusion of two boundary tachyon operators  because the Liouville 
 dressing  factors obey a  different fusion rule.  
 The correct interpretation comes from the 
  fusion rules \ACTAA\ involving one tachyon and one operator of 
  the boundary ground ring.
  Thus  the  operators  $A_+$  
     and  $A_-$ generating the boundary ground ring 
  have the   geometrical meaning of the operations of 
  adding or removing  lines.
   The action of these operators is encoded in the left and right
    boundary conditions           
     for the matter field:
  \eqn\AAbb{\matrix{
    A_+  &=&\  ^{s\pm { i\pi /b}}_{\chi+ { \pi /b} } \[A_+\]^{s}_{\chi } , 
&\qquad &
    A_-  &=&  ^{s \pm i\pi  b}_{\tilde  \chi-{ \pi }{b}}
      \[A_-\]^{s}_{\tilde \chi } \cr
      & & & & & & \cr
      \overline{A}_+  &=& \  ^{s\pm { i\pi /b}}_{\chi- { \pi /b} } \[ 
\overline{A}_+\]^{s}_{\chi } ,
      &\qquad & \overline{A}_- &=&
      \  ^{s \pm i \pi b}_{\tilde \chi+{ \pi }{b}}  \[ 
\overline{A}_-\]^{s}_{\tilde \chi } .}
    }

  \bigskip
  
\subsec{ A self-dual  $\hat A_\infty$ matrix chain}


\noindent
The   matrix chains associated with Dynkin diagrams \adem\    were 
   originally   defined in 
     terms of a set of $N\times 
N$   hermitian matrices  $\Phi_x$ 
   associated with the nodes $x$ of the Dynkin  diagram
and $N\times N$   complex matrices   $A_{\<\! x, x' \! \>}$  
   associated with the   links $\< \! x,x'\! \>$.
     All these  matrix chains  have a loop gas description and
     therefore  a direct Coulomb gas  interpretation.
     The complex matrices $A_{\<\! x, x'\! \> }$   represent the
     one type of degenerate boundary fields 
     (only electric or only magnetic ones).

  Each value of the central charge is realized by two such 
  matrix models,   related by duality.  
  For example   the theory with $b={p\over p+1}$  and  central charge 
  $c=1-{6\over p(p+1)}$  is realized either as an 
  $A_p$ chain  in the dilute phase   or as an $A_{p+1}$ chain 
   in the dense phase.
  The electric degenerate boundary  fields have a good description in 
  the $A_{p}$ matrix chain  while the
  magnetic degenerate boundary  fields have a good description in the 
  $A_{p+1}$  matrix chain.

 Remarkably,  each such pair of  matrix chains related by duality 
 can be  formulated as a single  {\it self-dual}   chain of complex matrices
 by the following simple prescription.
   The hermitian matrix variables $\Phi_x$ associated with the nodes of the 
   Dynkin diagram should be considered as  composite fields  
   $B_xB_{x}^{\dag}$ or   $B_x^{\dag}B_{x}$,  
   where $B_x$ are  $N\times N$ complex   matrices.
Here we will consider only the $\hat A_\infty$  chane, which
 provides the matrix model description  of  the theory \actg.
    The case of  the rational $ADE$ theories will be studied elsewhere \KP.

       The    partition function of  the self-dual $\hat A_\infty$   matrix chain   is 
       defined  in terms of the $N\times N$ complex matrices 
       $ A_{\<\! x, x' \! \>} \equiv A_x$ and $B_x$  $ (x\in \IZ)$:            %
\eqn\partfMM{
\eqalign{
\CZ =& 
\int \prod _{x\in \IZ} dA_{x} dA_{x} ^{\dag} dB_x dB_{x}^{\dag} \
\ e^{- \CS(A, A^{\dag}, B, B^{{\dag}} )}}}
where   the simplest action with linear 
potential for $AA^{\dag}$ and $BB^{\dag}$ is
     \eqn\matact{\eqalign{
      \CS(A, A^{\dag}, B, B^{\dag} ) = &
      \sum_{x\in\IZ}\Tr  \(
      2TA_{x}A^{\dag}_{x} +2  TB_x B^{\dag}_x-
      B_{x}B^{\dag} _{x} A_{x} A_{x} ^{\dag}
           -  B ^{\dag}_{x+1} B_{x +1}  A^{\dag}_{x} A _{x}   
      \)   . \cr
     }
     }
     (Of course this   integral  has only formal meaning
     because of the infinite volume of the chain. 
       It can be regularized 
        as the limit $n\to\infty$ of the  periodic chain $\hat A_n$.)
        
   Alternatively one  can write the partition function  as an integral over the 
       positive  definite
 hermitian    matrices   
  \eqn\PhiPhid{
  \Phi_x = B_{x}B^{\dag}_{x }  ,\qquad  \tilde \Phi_x  =  A_{x} 
A^{\dag}_{x}
         }
         and the unitary matrices $U_x$.
 Let us denote by $[d\Phi]_{>0}$ the flat measure over the positive 
definite 
 hermitian matrices and by 
 $[dU]_{_{SU(N)}}$ the Haar measure on the group $U(N)$.
 Then the integral \partfMM\ can be written as
 \eqn\partPHI{
 \CZ 
 =\int \prod_{x\in\IZ} \  [d\Phi_x]_{_{>0}}[ d\tilde\Phi_x  ]_{_{>0}}
 [DU_x] _{_{SU(N)}}\ e^{-\CS (\Phi, \tilde\Phi, U)}
 }
 \eqn\actPHI{\eqalign{
 \CS (\Phi, \tilde\Phi, U)=
 \sum_{x\in\IZ}
 \Tr \( 2T \Phi_x   +2T \tilde \Phi_x-
     \Phi_x\tilde \Phi_x -  U\Phi_x U^{-1} \tilde \Phi_{x +1}\)
 }
     }
        The integration over the  unitary group can be done 
explicitly
      and the  partition function can be written as an integral
       with  respect to the eigenvalues  $\phi_{x,j}$ and $\tilde 
\phi_{x,j}$ of $\Phi_c$ and $\tilde \Phi_x$ respectively.
        Using the Harish Chandra-Itzykon-Zuber formula
         we  write \partPHI\  in the form   
               \eqn\mteig{
     \CZ = 
      \int \limits_{0} ^\infty \prod _{x , j}\  d \phi_{x,j }
      \ d \tilde \phi_{x,j }\ e^{-2T\phi_{x,j}-2 T\tilde  \phi_{x,j}}\ 
    \det _{jk} e^{ \phi_{x,j }\tilde \phi_{x,k }}
     \det _{jk} e^{ \phi_{x,j }\tilde \phi_{x+1,k }}.
        }

     We are  interested  in the 't Hooft limit  
     $N\to\infty$ with $\kappa\equiv N/T^2 $ fixed.
      Then the    ground state of the matrix model  
    is characterized by the classical  spectral densities
   $\rho_x(\phi)$ and $\tilde \rho_x(\tilde\phi)$
   of the hermitian matrices \PhiPhid.
 The 't Hooft limit exists for 
    $\kappa < \kappa_c$, where $\kappa_c$ is the critical coupling
    where  the planar graph expansion  diverges.
 One can show that in this case   $\rho_x(\phi) $ and $\tilde \rho_x(\tilde\phi) $ 
  vanish for $\phi_x, \tilde \phi_x  \ge T$, up to exponentially small in $N$ terms. 

  We will be  interested only in the scaling limit, which  describes the 
  infinitesimal vicinity of the critical point. 
  It is technically  simpler  to  treat  the parameters  
  $N$ and $T$ as  large  but  finite.  Then for  given  $T$ 
  the critical point is achieved at 
   $N_c= \kappa_c T^2$.  The value $N_c$   plays the role of a 
   cutoff parameter, which we eventually sent at infinity, 
    while   the   cosmological constant  is   $\mu\sim N_c-N$.

   The matrix chain \matact\   describes    the
     particular case $b=1$ of the world sheet  theory \actg, 
     when the theory is unitary.  A non-zero background charge 
     $e_0=1/b-b$ can be introduced   by 
     imposing by hand    a $x$-dependent background.  
     Then the  spectral densities are  solutions of the saddle point 
     equations under the constraints
     \eqn\timedpd{
     \rho_x(\phi) = e^{i\pi \e_0x /b} \ \rho(\phi), \quad
       \tilde \rho_x(\phi) = e^{-i\pi \e_0 bx} \ \tilde \rho(\phi).
       }

              \bigskip

    \noindent
        $\circ$\ {\it Relation to the standard $\hat A_\infty $ chain \adem}

         \noindent  One  can  formally  integrate   
         with respect to the  $\tilde\phi$-variables  in \mteig\ 
         neglecting the contributions of infinity. In this way one neglects 
     exponentially small in $N$ terms, which  do not influence the genus expansion.  The result of the integration is
      the partition function of the  hermitian 
$\hat A_\infty$ matrix chain    \adem\  
     \eqn\mahat{
     \CZ_{\hat A_\infty}=\int\limits _{0}^\infty \prod _{x,  j}\  d 
\phi_{x,j }  e^{-2T\phi_{x,j}}\ 
      {\prod _{x} } 
    {\prod _{j\ne k} \( \phi_{x,j }- \phi_{x,k }\)\over 
     \prod _{ j, k}  \(2T- \phi_{x,j }  - \phi_{x+1,k }\)
     }.}
  By  integrating first with respect to the $\phi$-variables one 
obtains the  same eigenvalue  integral for the dual field
    $\tilde \Phi_x$:
         \eqn\mahad{
     \CZ_{\hat A_\infty}=\int\limits _{0}^\infty \prod _{x,  j}\  d 
\tilde \phi_{x,j }  e^{-2T\tilde \phi_{x,j}}\ 
      {\prod _{x} } 
    {\prod _{j\ne k} \(\tilde  \phi_{x,j }- \tilde \phi_{x,k }\)\over 
     \prod _{ j, k}  \(2T-\tilde  \phi_{x,j }  -\tilde  \phi_{x+1,k }\)
     }.}    
 The saddle point    equations for  these integrals lead to 
 spectral densities
$\rho_x(\phi)$ and $\tilde \rho_x(\tilde \phi)$ that 
vanish for $\phi, \tilde\phi>T$, which justifies the  fact that we 
neglected the 
contributions  at infinity.

     \bigskip
     
  \noindent
        $\circ$\ {\it Relation to  Matrix Quantum Mechanics}
     
\noindent
It is known \Kstau\  that the hermitian 
$\hat A_\infty$ matrix chain  \mahat \ and MQM 
share the same $x$-space  correlation functions of 
on-shell closed  string tachyons. 
Thus the $\hat A_\infty$ chain is expected to  yield  a discretization of
 the singlet sector of 
 MQM  at euclidean time distance  
 $\b=\pi/2$ in the normalization in which 
     the self-dual distance is $\b=2\pi$.
    Let us see  that this is indeed the case.

      The  singlet sector of  MQM is equivalent to 
     a system of $N$ nonrelativistic fermions in upside-down 
     oscillator potential $U(y)= -\hf y^2$,
     representing the $N$ eigenvalues of  a  hermitian matrix   
     variable 
     (see the review \KlebanovMQM\ and the references 
     to the original papers therein). 
    The evolution of each eigenvalue $y$ 
         in MQM is described by a one-particle 
         Hamiltonian $H_0=- \hf(\p_y^2+y^2)$. 
       The  heat kernel 
         for this   Hamiltonian     is given by 
\eqn\kernb{K_{\b}(y,\tilde y)\equiv  \langle y| e^{ -\b H_0} |\tilde y\rangle  
                = {1\over \sqrt{2\pi \sin \b}}\ \exp\(
{ 2y\tilde y - \cos \b  (y^2 +\tilde y^2) \over 2\sin \b}\).
}
Therefore the    Laplace kernels entering 
 the    determinants in \mteig\ can  be viewed, after a linear change of variables 
 $$y = \phi  - T, \  \ \ \tilde y = \tilde \phi - T,$$
 as
   evolution   kernels   \kernb\ with      $\b= \hf\pi$.
   Let us denote by $\CH=\bigoplus_{k=1}^N H_0$   the  MQM 
 Hamiltonian in the singlet representation and by $\CP_{T}$ 
 the projector restricting the integration to the 
 semi-infinite  interval $[-T, \infty]$:
 $$\Theta_{_T}= \prod_{i=1}^N \theta(y_i -T).$$ 
 Then the   partition function  \mteig\  for a periodic chain
 $\hat A_n$  can be written as  the trace
   \eqn\hatan{\eqalign{
\CZ_{\hat A_n}&=  \Tr 
\(\Theta_{_{T} }\ e^{-\hf \pi \CH} \)^{2n}.
}}
The projector $\Theta_{_{T} }$ has the effect of an infinite potential wall 
placed at  distance  $T$ from the origin, which stabilizes the inverse oscillator potential.  For given $T$ the Fermi level grows with the 
number of the eigenvalues $N$ and  at some critical $N=N_c$ 
reaches the top of the potential. The cosmological constant  
$\mu\sim N_c - N$ measures the deviation from the critical point.

         \subsec{Disc loop amplitudes  in the matrix chain  }

         \noindent
          The classical background of the matrix chain is determined by the
           (unnormalized) disc loop amplitudes
         \eqn\defWx{W_x(z) = \<{ \Tr } {1\over z- B_x B_x ^{\dag}  } \>
         =\int {d\phi \rho_x(\phi)\over z-\phi}}
         and 
             \eqn\defWxd{  
          \tilde W_x(\tilde z) 
           = \< {\Tr} {1\over \tilde z- A_x A_x ^{\dag}  }\>
           =\int {d\phi \tilde \rho_x(\tilde \phi)\over \tilde z-\phi}.
           }
           
           The  loop amplitude $W_x(z)$
            represents a meromorphic function in the spectral parameter $z$ 
with a cut along the support  $[0, a]$ of the eigenvalue distribution, where 
$0<a\le T$.  It  is determined from     the saddle-point equation 
for the integral \mahat, 
which can be  written in the form of a
boundary condition along the cut:
   \eqn\loopeqZ{
     W_x(z+i0) +W_x(z-i0)+   W_{x+1}(2T-z)+W_{x-1}(2T-z)=2 T
     \qquad (0<z<a).
     }
The dual loop amplitude  $\tilde W_x(\tilde z) $ has the same  properties
with $z, a$ replaced by $\tilde z, \tilde a$.
 
  The true ground state  does not depend on $x$ and  describes the unitary 
     theory with matter central charge $c=1$.
   The  non-unitary theory with   background momentum $e_0$
    is obtained by 
     imposing by hands the following constraints on the 
     loop amplitudes:\foot{It is actually sufficient to impose  only the second
     condition; then the first one follows  from the correspondence 
     $   W   \leftrightarrow  \tilde z, \ z \leftrightarrow\tilde W $.}   
      \eqn\Wxxd{W_x(z) = e^{i  e_0x/b }  \ W (  z),
            \qquad
      \tilde W_x(\tilde z) = e^{-i  e_0xb }  \ \tilde W (\tilde   z).} 
      Then the  equation \loopeqZ\    becomes
        \eqn\lpqz{
     W(z+i0) +W( z-i0)-  2\cos  (\pi /b^2)\
      W (2T - z)
     =2T \qquad (  0< z<  a) .
     }
This equation is readily solved in 
 the scaling limit, when   ${N-N_c\over N_c}$ and ${|z-T|\over T}$ are 
 considered small. 
Then    the   branch point at $z=T$  can be resolved by the parametrization 
  \eqn\ztos{
     z - T=   M \cosh (\pi  sb) , \qquad M= T-a,
     }
and    \lpqz\    becomes  a difference equation
   \eqn\loopeqs{
  W(s+ {ib} )  + W(s-{ib} ) - 2\cosh {\pi /b^2}\
  W(s )= 2T.
 }
 From here we find 
    \eqn\Wsolu{
  W(s)  =c_1 T + 
  c_2  T^{1-b^2} \  M^{1/b^2}\cosh (\pi s/b).
  }
   where $c_1$ and $c_2$ are $b$-dependent  constants.
    The solution reproduces, up to a linear transformation,  the loop amplitude 
       (D.16) in the  world sheet CFT. 
  
In the following we will retain only the scaling parts of $z$ and $W$,
assuming   a normalization that matches with the result from the continuum CFT:
\eqn\zW{
z=M\cosh(\pi b s), \quad W=\  {M^{1/b^2}\over {1\over b} \cos({\pi\over b^2})}
\cosh (\pi s/ b).
} 
 In this normalization the  eigenvalue density 
    is given by   
    \eqn\ruos{
     \rho(z)  ={ W(s-\frac{i}{b} ) -  W(s+\frac{i}{b} ) \over 2\pi i}=
    bM^{1/b^2} \sinh (\pi s/b) .
    }
    Proceeding in the same way we find the scaling part of the
 dual  loop  amplitude \defWxd
 \eqn\tilzW{
 \tilde z=\tilde M\cosh(\pi b /s), 
 \quad \tilde W=\  {\tilde M^b \over { b} \cos({\pi b^2})}
\cosh (\pi sb),
   }
 where $\tilde M= T-\tilde a$.

     The loop amplitudes  \zW\  and  \tilzW\ are related by a duality 
 transformation 
 \eqn\WWdual{
  \tilde W= -{1\over b\sin\pi b^2} \ z, \qquad 
 \tilde z = - \frac{1}{b} \sin (\frac{\pi}{b^2})  \  W.
 } 
  Intuitively,
    the duality in Liouville gravity is a kind of electric-magnetic duality
 on the world sheet, whose target-space equivalent is a Fourier transformation
 for the eigenvalues of the random matrix.

  In the planar limit the matrix model is described by a classical Hamiltonian system in which the spectral parameter $z$ and 
   the resolvent $W$ 
  form a pair canonically conjugated variables: the coordinate and the momentum.  Then the  duality  make sense of a canonical transformation exchanging  the momentum and the coordinate. 
  The  meromorphic functions    $W(z)$  and  $\tilde W(\tilde z)$  
   give two parametrizations of the  complexified classical trajectories
  of the Hamiltonian system.   
  
  The  physical sheets of the Riemann surfaces of $W(z)$ 
  and $\tilde W(\tilde z)$ can be viewed, together with the identification 
  \WWdual, as two charts   defining a complex curve with the topology of a 
  sphere with two punctures. 
In the rational models of quantum gravity the Riemann surface is
   given by an algebraic curve.    In the case of generic $b^2$ the
   equation of the algebraic curve can be written as a {\it functional }
   equation 
   \eqn\WZ{
 W^2(z) + W^2(-z)-2   \cos (\frac{\pi }{b^2})\ W(z) W(-z) = b^2 M^{2/b^2}
 }
or its dual
    \eqn\WZd{
\tilde  W^2(\tilde z) + \tilde W^2(-\tilde z)-
2   \cos (\pi b^2)\ \tilde W(\tilde z)
 \tilde W(-\tilde z) = \frac{1}{b^2} \ \tilde M^{2b^2}.
 }    
 These equations   are particular case of
  more general  functional equations that hold
  before the  scaling limit  and allow to express the 
  positions $a$ and $\tilde a$ of the branch points 
  in terms of  $T$ and $N$.
   In particular, the non-scaled  functional equation for
    $\tilde W(\tilde z)$, whose   derivation   presented in Appendix E,
     implies  the scaling $\tilde M^{2b^2} \sim\L$.

             \subsec{Planar graph expansion and  
             mapping to the world-sheet  CFT}

             \noindent
     The  perturbative expansion  of any observable 
      in the theory  with action \matact\ 
      is   a sum of  planar  graphs     composed 
      from  the vertices 
     $$\matrix{  & ^{A^{\dag}_x }&\cr
^{A_x }&{  \epsfxsize=29pt
          \epsfbox{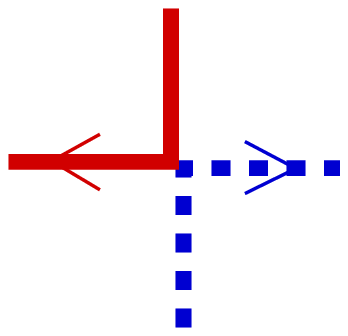}  }&  ^{B_{x}}\cr
          & ^{B^{\dag} _{x} }&}\qquad {\rm and } \qquad 
            \matrix{ & ^{A_x} &\cr
^{{A_x} ^{\dag}}&{  \epsfxsize=29pt
          \epsfbox{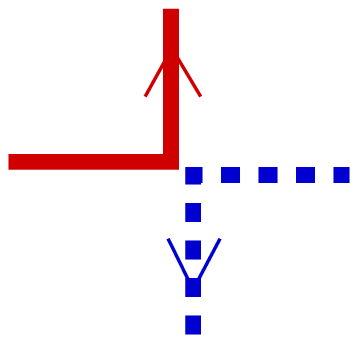}  }&  ^{B^{\dag} _{x+1}}\cr
          & ^{{B_{x+1} } }&},
          $$       
where the  continuous and dashed oriented  lines  represent
respectively  the 
  propagators of the $A$ and $B$-matrices.  The planar graphs can 
  be  considered as discretized  world sheets embedded in $\IZ$.
   The    loops made of the  $A_x$-propagators  can be viewed as
    domain walls
   separating  domains with  time coordinates  $x$ and $x+1$.

We are interested in   the
           disc amplitudes representing  various   boundary correlation 
           functions of tachyon operators.    The   simplest   ones 
            are expectation values   of the resolvents 
            \defWx\ and \defWxd, which 
            are the equal to the boundary one-point functions in the 
            world sheet CFT  with Neumann and Dirichlet boundary
             conditions\foot{Strictly speaking, the    Neuman boundary condition
             is generated by the resolvent of the operator $(A+A^{\dag})^2$.
             However   it has the same scaling limit as  the resolvent of
             $A^{\dag}$. }.

          In the planar  limit   $W_x$ 
           is a sum of   planar graphs
          with the topology of a disc,
           made by $A$ and $B$ type  lines. The $A$-lines can only  form  
           loops while the $B$-lines can either close in a loop or  have its both ends  among the external legs. 
          The condition \Wxxd\ is equivalent to 
              associating     $x$-dependent phase factors 
       with the vertices and  faces of the planar 
   graph as explained in section 2 of \Idis.
         For example, the  graph shown in  fig.5 gives a  particular 
        discretization of the   world sheet
           with three domain walls    shown in fig.4.

            \vskip 10pt
          \centerline{ \epsfxsize=130pt
           \epsfbox{  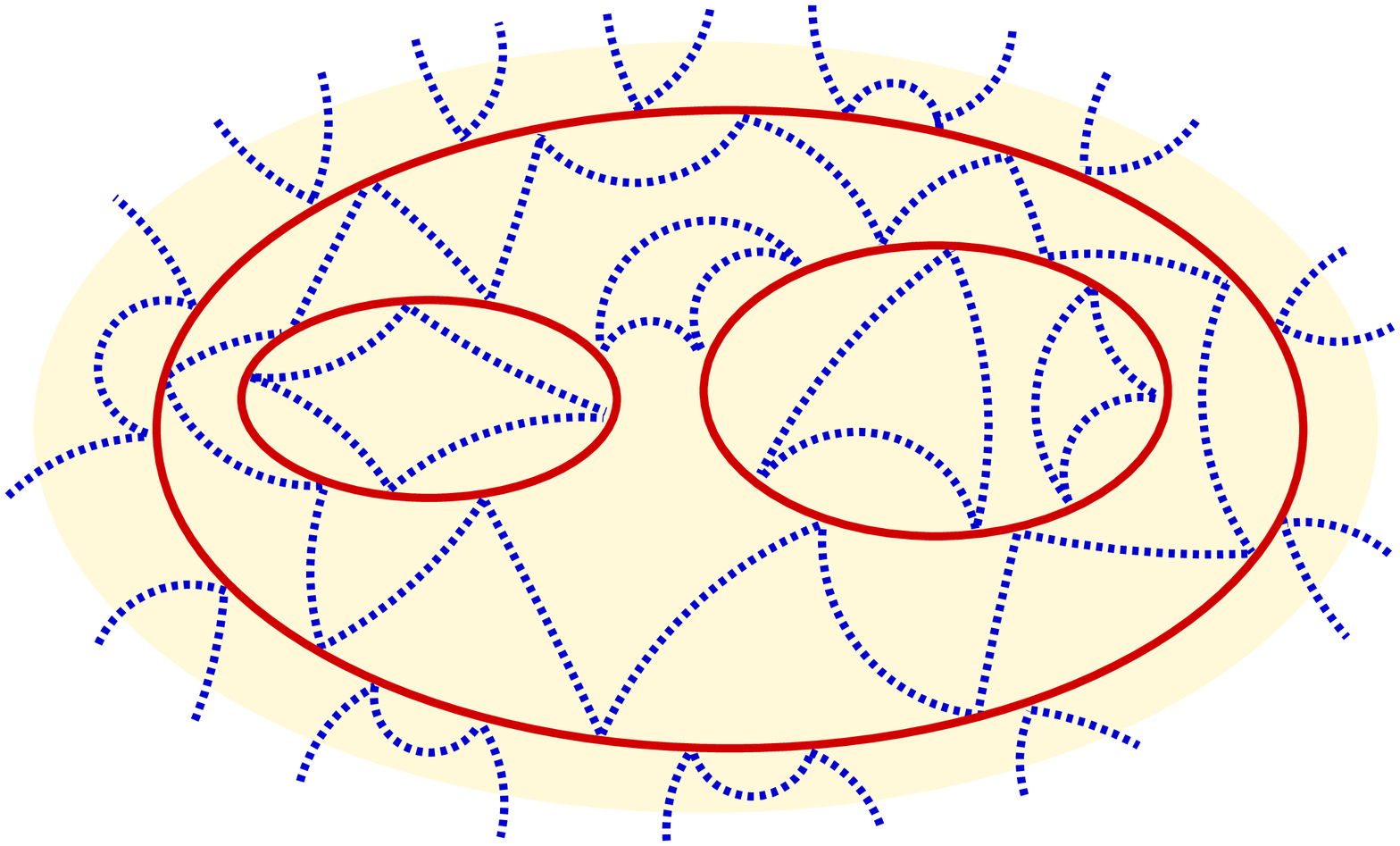  } \hskip 50pt 
          \epsfxsize=125pt
           \epsfbox{  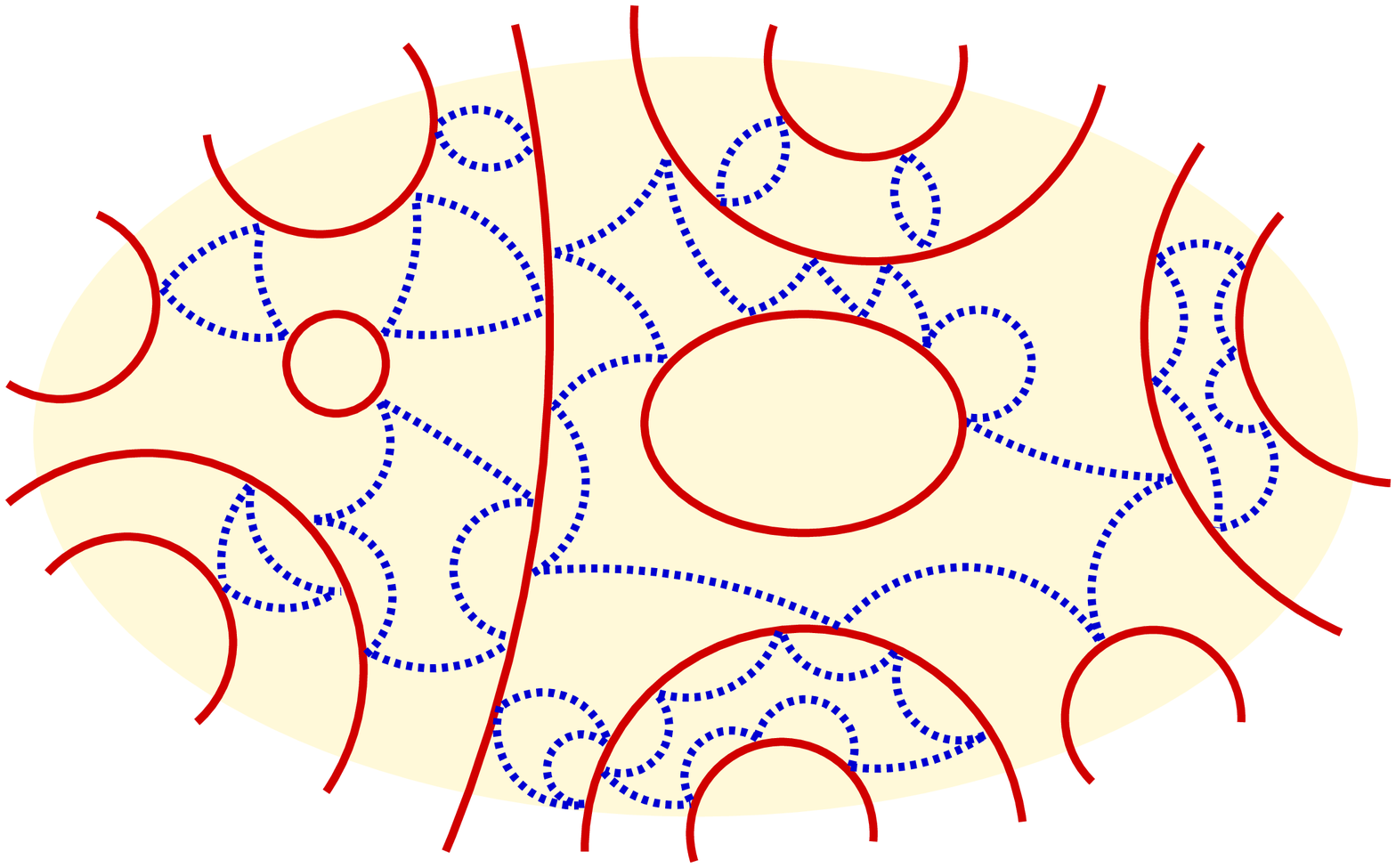  }  }
           \vskip 5pt

           \centerline{\ninepoint Fig.5 a,b:  Planar graphs contributing to  
           the resolvents of $B_x B_x ^{\dag}  $ and $ A_x A_x^{\dag} $.}
            
            \vskip  10pt

        \noindent 
         
         The  dual  amplitude $\tilde W_x$  is a  sum of planar graphs 
          such that  the  $B$-lines are always closed and the $A$-lines either close or  connect two external legs.             
            In this way the complex matrix chain allows to define 
         both the Dirichlet and Neumann boundary conditions for the loop gas in the sense of \KKloop \ and  \KPS.   
         Exchanging the $A$ and $B$ lines 
         is  equivalent to the duality transformation, which exchanges 
         Dirichlet and Neumann boundary conditions.
         In the continuum limit  the two amplitudes are given,
         up to  rescaling and subtraction of a constant, by the expressions
         (D.16) and (D.17).

     The boundary ground ring is represented in the 
     matrix model as the algebra of the polynomials of the 
     $A$ and $B$ matrix variables. For the generators 
     of the algebra the correspondence is
      \eqn\grrgM{ 
    \matrix{   A_+\to
   A_x   & \qquad A_-     \to
  B_x\cr
  &\qquad \cr
    \overline A_+ \to 
    {A}^{{\dag} }&\qquad 
   \overline{ A}_-      \to  B^{\dag }.
  \cr}
  }
     
     The   boundary segments   with  boundary condition 
  $(\chi, s)$ or $(\tilde \chi, s)$ are represented by the resolvents
  of the fields $\Phi_x= B_x B_x^{{\dag}}$ and 
  $\tilde \Phi_x=A_x A_x^{{\dag}}$ correspondingly:
  \eqn\CFTMb{\eqalign{
  ]_\chi^s[& \quad \to\quad  {1\over z(s) -\Phi_x}\cr
   ]_{\tilde \chi}  ^s[& \quad \to\quad {1\over z(s) - \tilde \Phi_x}
   .}}
        The degenerate boundary operators  $\CB_{n,1}$
     and $\CB_{1,n}$ 
     are created by  inserting  in the trace   polynomials 
     of $A$ and $B$ matrices according to the following dictionary:
    \eqn\CFTMM{
    \matrix{  _{\chi +n \pi/b} [\CB_{n,1}] _{\chi}\to
  A_{x+n}...  A_x&\qquad 
  _{\tilde \chi +n \pi b} [ \CB_{1,n}] _{\chi}\to
  B_{x+n}...B_x\cr
  &\qquad \cr
   _{\chi - n \pi/b} [ {\CB_{-n,1}}] _{\chi}\to 
   {A_{x-n}^{\dag} ...  {A_x}^{\dag} }&\qquad 
  _{\tilde \chi - n \pi/b} [ {\CB_{1,-n}}] _{\chi}\to 
  B_{x-n}^{\dag}...B_x^{\dag} .
  \cr}
  }
  
       For example, the boundary two-point function $D_P(s,s')$ 
   for $P= \hf e_0 + {n \over 2b} $ 
    corresponds to the following
   expectation value in the matrix model (Fig.4):
   \eqn\corfAn{\eqalign{
  D_{  {e_0\over 2}  -{n\over 2b}  } (s, s')&= 
   \< ^{\ \ \ \ s'}_{\chi + n {\pi\over b} }[\CB_{ n+1,1}] ^s _\chi 
[\CB_{-n-1,-1}]
  _{ \chi + n {\pi\over b} }^{s'} \>_{\rm disc}
   \cr &\sim    \<  \Tr\   {A_{x+n}^{\dag} }... {A_{x}^{\dag} }
    \ {1\over z(s) - \Phi_x}
    A_{x}...  A_{x+n}
    \  {1\over   z (s')- \Phi_{x +n} }\>.
   }
   }

   \noindent
 This  correlation function was calculated from the matrix model 
side using the loop gas technique  \refs{\Ibliou, \KPS} 
   and the result is in agreement with the prediction of the 
Liouville theory.
   Another example is the  amplitude of two excited 
   twist operators with momenta 
   $\pm P_{n+\hf, 0}= \pm {(n+\hf)/2b} $, which  
    interpolate between
    Dirichlet and Neumann boundary conditions: 
   \eqn\Ttwists{\eqalign{
   D_{{1\over 2b} (n+\hf)}(s, s')&=
   \<  ^ {s'}_{\tilde \chi  }[\CB_{{n+1/2\over 2b} }] ^s _\chi 
[\CB_{{n+1/2\over 2b} }]
  _{\tilde  \chi  }^{s'} \>_{\rm disc}\cr
  &\sim  
   \< \Tr\  {A_{x+n}^{\dag} }... {A_{x}^{\dag} }\
    {1\over  z(s)- \Phi _x} \   A_{x}...  A_{x+n}\
    {1\over \tilde z(s') - \tilde  \Phi_x} 
  \>.
   }
   }
    This correlation functions  has been calculated (for the case 
$n=0$)  in 
    \KKloop\  and given a Liouville CFT interpretation in  
\refs{\Ibliou,\KPS}.
 
 %
%
%
%
%
%
    
    \newsec{Concluding remarks}

    \noindent
 1.   The   ring  relations  and the 
 functional equations   
 derived   in this paper 
 can be generalized to the case of  any matter CFT with $c<1$.
 This  can be achieved by 
 replacing the $U(1)$ fusion rules  for the gaussian field 
 with  the fusion rules in the corresponding 
 boundary CFT.    
  
   \smallskip
   \smallskip
 
 \noindent
 2.  
   In  the  $ADE$ 
   rational   string theories, 
the matrix model realization 
    of the boundary ground ring   
    can be achieved by reformulating the 
      corresponding  
    $ADE$ 
      matrix  chain  
in terms of 
 $A$-type  and $B$-type  complex matrices.

  Consider for example the simplest case of  a theory of the 
$A$-series
  with central charge $c= 1- {6\over p(p+1)}$, which corresponds to   
$b^2 = {p\over p+1}$.  The corresponding string theory is described   
by the   critical point of the 
  $A_p$  matrix model  or by  the tricritical point of the 
$A_{p-1}$   matrix model.
 Both  models can be incorporated in a single matrix model 
 of the type \matact\ by replacing the Dynkin graph of $\hat A_\infty$
 with that of $A_p$. The new  model is defined  in terms of $p-1$ 
$A$-matrices labeled by the links of the Dynkin graph and $p$ 
$B$-matrices  labeled by the points of the graph.

 \smallskip \smallskip
 
  \noindent
  3. It is interesting that  Dijkgraaf and Vafa  \DV\ proposed a 
similar matrix 
  representation for the {\it chiral} ring for the 
  $c=1$ string theory  compactified at the self-dual radius.
     Since the boundary operators can be considered as operators in a 
chiral
      CFT \refs{\ReSch,\ValyaJB}, there might be a connection between 
the  Witten's chiral ring and the boundary ground ring.

              \smallskip \smallskip
              
\noindent
4. The action of the boundary ground 
 ring   closes  on the  discrete subset   $s=  i {m\over b} + inb,
$ $m,n\in \IZ$. 
This subset is relevant for the  
ZZ brane boundary states \ZZPseudo, as discussed in 
\refs{\MartinecKA, \newhat, \TeschnerQK,  \SeibergS}.
One might ask oneself whether  in the
ZZ brane there is an analog of the boundary correlation functions 
in the   FZZT brane.

\noindent

   \smallskip\smallskip\smallskip
 \bigskip
\noindent
{\bf  Acknowledgments}
\smallskip

\noindent
The author thanks  B. Ponsot, 
V. Schomerus, D. Serban,  J. Teschner, Al. Zamolodchikov   and 
especially 
V.  Petkova for valuable discussions.
 This research   is supported in part by the 
 European network  EUCLID, HPRN-CT-2002-00325.

 \appendix{A}{ Bulk and boundary  reflection  amplitudes}

 \noindent
The  closed and open string tachyons of same momenta 
but opposite chiralities 
 are related by the bulk and 
Liouville boundary reflection amplitudes \FZZb
\eqn\reflBul{
\CV_P{(+ )}= S_P^{(+-)} \CV_P^{(-)}, \qquad
\CV_P{(-)}= S_P^{(-+)} \CV_P^{(+)}
}
\eqn\reflB{
  [\CB^{(+)}_P]^{s _1s _2} =  D_P^{(+-)}  (s _1, s _2) 
\ [\CB^{(-)}_P]^{s _1s _2}, \quad
 [\CB^{(-)}_P]^{s _1s _2} =   D_P^{(-+)}( s _1,  s _2) 
\ [\CB^{(-)}_P]^{s _1s _2}.
}
In  the normalizations \newnor\ and   \normB 

\eqn\Spm{S_P^{(+-)} =  - {1\over b^2}\ \L^{P/b}, \qquad 
S_P^{(-+)} =  - {  b^2}\ \L^{-P/b}}

   \eqn\DsubP{\eqalign{
 &D_P^{(+-)}( s _1,s _2) =
  {1\over b}  { \L^{P/b}  \over S_b(2P+b)
    }\ 
   {S_b\({Q\over 2} +P - i {s _1+s _2\over 2}\)
  \ S_b\({Q\over 2} +P - i {s _1-s _2\over 2}\)\over 
  S_b\({Q\over 2} -P - i {s _1+s _2\over 2 }\)
  \ S_b\({Q\over 2} -P - i {s _1-s _2\over 2 }\)}
}
}
where $\G_b(x)$ is the Double Gamma function (see \FZZb)
and $S_b(x)= \G_b(x)/\G_b(Q-x).$
 The    amplitude  $D_P^{(+-)} = (D_P^{(-+)} )^{-1}$   has
 the    symmetries
 \eqn\REFLD{
 D_P^{(+-)}D_{-P}^{(+-)} = {1\over b^2}  {\sin {2\pi\over b}  P
 \over\sin 2\pi b P}, \qquad
 D_P^{(-+)} D_{-P}^{(-+)}= b^2 {\sin 2\pi b P
 \over\sin {2\pi\over b}  P}
}
and 
\eqn\REFB{
D_{-P}^{(-+)} (s _1,s _2) =
\tilde D^{(+-)} _{P}(   s _1,   s _2),
}
where the function 
$\tilde D_P^{(+-)} (s _1,s _2)$ is defined by \DsubP\
with $b$  replaced by   $\tilde b = 1/b$.
 The reflection amplitudes  for  the degenerate momenta 
  $ P_{mn} = \hf(m/b - nb)$  are given by rational functions of
   $z=M\cosh (\pi b s )$ and $\tilde z = M^{1/b^2}\cosh(\pi s/b)$.
   For example,
  \eqn\exmples{\matrix{
  D^{-+}_0 &  =&  b S(b)= b^ 2	& \qquad \quad &
  D^{+-}_0  &=&  {1\over b }S(1/b)= {1\over b^ 2}
   \cr
  D^{+-}_{b/2}&=&  {b^{-2}   \over  \sin( \pi b^2)}( z_1+ z_2)
  & \qquad \quad &
   D^{-+}_{-1/2b }&= &{b^2 \over \sin(\pi/b^2)}  
  (\tilde z_1+ \tilde z_2)\cr
    D^{-+}_{-1/2b + b/2}&=&- {b^2\sin( \pi b^2)\over \sin(\pi/b^2)}  
  {\tilde z_1- \tilde z_2\over z_1-z_2}& \qquad \quad &
    D^{+- }_{1/2b - b/2}&=& 
 - {\tilde z_1- \tilde z_2\over z_1-z_2}.
  }
  }

 \appendix{B}{Recurrence relations for the  closed string  tachyon 
 amplitudes
 }

 \noindent
Using  the ring relations  {\actaa} and  {\actaba}
 one   can obtain a set of recurrence  equations 
 for the   correlation functions of the bulk tachyons
 \eqn\defG{
 \eqalign{
 &G(P_1,...,P_n|P_{n+1},...,P_{n+m})=\cr &
 \< \[\prod_{k=1}^{n-1}
 \int 
  \CV^{(-)}_{P_k}\]
  \
 {\rm\bf c\bar c}  \CV^{(-)}_{P_n}
   (0)  {\rm\bf c\bar c}\CV^{(+)}_{P_{n+1}}(1)  
\[
  \prod_{j=n+2}^{n+m-1}
  \int 
\CV^{(+)}_{P_{j}}
 \]
  {\rm\bf c\bar c} \CV^{(+)}_{P_{n+m}}(\infty)
\>
 }
 }
 which generalize the recurrence relations for 
 the resonance amplitudes obtained in \bershkut.
 Here we assume that the neutrality condition 
 is satisfied
 \eqn\neutrality{\sum_{k=1}^{n+m} \( e_0- P_k\) =2e_0
 .}
 The auxiliary function 
 \eqn\auxF{\eqalign{
 &F(x,\bar x |P_1,...,P_n|P_{n+1},...,P_{n+m})=\cr &
 \< \[ \prod_{k=1}^{n-1}\int
  \CV^{(-)}_{P_k} 
  \]
   {\rm\bf c\bar c} \CV^{(-)}_{P_n+b}
   (0)\ a_-(x,\bar x)\  {\rm\bf c\bar c} \CV^{(+)}_{P_{n+1}}(1)
  \[\prod_{j=n+2}^{m+n-1}\int  \CV^{(+)}_{P_{j}}
 \]  {\rm\bf c\bar c} \CV^{(+)}_{P_{n+m}}(\infty)  \>
 }
 }
does not depend on $x$ and $\bar x$.  This can be proved by   using
$\p_x a_-=\{Q_{\rm BRST}, {\rm \bf b}_{-1} a_-\}$
and deforming the contour, commuting $Q_{\rm BRST}$ with the other 
operators
in \auxF \ \bershkut. Therefore one can
evaluate this function at $x=0$ and $x=1$ by
using the fusion rules  \actaa \ or  \aminus \ and  \actaba. 
As a result one obtains the recurrence relation

      \eqn\recCL{
       \eqalign{
      &G(P_1,...,P_n|P_{n+1},...,P_{n+m})    =   \L \   
     G(P_1,...,P_n+b|P_{n+1}-b,...,P_{n+m})-\cr &
     -\sum_{j=1}^{m-1} 
     G(P_1,...,P_n+b |P_{n+2},
     ... , P_{n+j} +P_{n+1} -\frac{1}{b},  ...,P_{n+m})
     .}
     }
 Similarly, by inserting   $a_+$    
 we get  the dual recurrence relation
    \eqn\recCLd{
     \eqalign{
      & G(P_1,...,P_n|P_{n+1},...,P_{n+m})    = 
       \tilde  \L \   
     G(P_1,...,P_n+\frac{1}{b}|P_{n+1}-\frac{1}{b} ,...,P_{n+m})-\cr &
      -
      \sum_{k=1}^{n-1} 
      G(P_1,...,P_k +P_n+b, ..., P_{n-1}|P_{n+1},...,P_{n+m}).}     }
          Note that the three-point function  $m+n=3$ coincides with the 
  corresponding  Liouville  three-point function.
  The latter has been evaluated   for generic momenta  ({\it i.e.}
  without imposing  the neutrality condition \neutrality\ by using a 
very similar argument   in   \Teschner.

        \appendix{C}{Recurrence relations for the boundary 
correlation functions
        with pure Neumann boundary conditions.}

        \noindent
    The fusion rules above   lead to a system of  recursion relations 
for the disc correlation functions with arbitrary number of  boundary 
states
    \eqn\defWnoz{\eqalign{
    &
    W(P_1,..., P_n|P_{n+1}, ..., P_{n+m})=\cr
    &
     \< \[ \prod_{k=1}^{n-1} \int   \CB  _{P_k}^{(-)} \]
        {\rm c} \CB  _{P_n}^{(-)}(0)  \  {\rm c}
   \CB  _{P_{n+1}}^{(+)}(1) 
       \[ \prod_{j=n+2}^{n+m-1} \int    \CB  _{P_j}^{(+)} \]
   {\rm c} \CB  _{P_{n+m}}^{(+)}(\infty)\>.
   }
   }
   Here three of the integrations are canceled by the volume of  the 
   global  $SL(2,\IR)$ conformal symmetry of the upper half plane.
   Consider  the auxiliary function
     \eqn\Fopen{\eqalign{
     &F(x|P_1, ..., P_n| P_{n+1}, ..., P_{n+m})
     =\cr
     & \< \[ \prod_{k=1}^{n-1} \int  \CB  _{P_k}^{(-)} \]
     {\rm c}   \CB  _{P_n+\frac{b}{2} }^{(-)}(0) A_- (x) \
 {\rm c}    \CB  _{P_{n+1}}^{(+)}(1) 
  \[ \prod_{j=n+2}^{n+m-1} \int    \CB  _{P_j}^{(+)} \]
    {\rm c} \CB  _{P_{n+m}}^{(+)}(\infty)\>.
}
    }
    Since 
    \eqn\brstopen{
    \p_x  F= \{ Q_{\rm BRST}, {\rm \bf b}_{-1} a_-\}
    }
  the function $F$ does not actually depend on $x$ and one can 
  calculate it in two different ways by taking the limits $x\to 0$ 
and $x\to 1$
  and using \ACTAA\ and \nonAr\
  (see \bershkut\ for a discussion concerning   possible boundary 
terms). 
  Note that  the boundary operators are ordered and the fusion can be 
performed only with the operators  $ \CB  _{P_{n}}^{(-)} $ and 
  $  \CB  _{P_{n+1}}^{(+)}$. 
  This leads to the recurrence relation
      \eqn\reqBd{  W(P_1,..., P_n|P_{n+1}, ..., P_{n+m})
      =-
        {W(P_1,..., P_n+\frac{b}{2}|P_{n+1}+ P_{n+2}- \frac{1}{2b},
        P_{n+3}, ..., P_{n+m})\over 
      \sin (2\pi b P_{n+1})}
      }
 or, after shifting the momenta,
    \eqn\reqBd{  W(P_1,..., P_n-\frac{b}{2}
    |P_{n+1}+\frac{1}{2b}, ..., P_{n+m})
      =
        {W(P_1,..., P_n|P_{n+1}+ P_{n+2},
        P_{n+3}, ..., P_{n+m})\over 
      \sin (2\pi b P_{n+1})}
      }
    and similarly for $A_+$, which can be easily solved:
    $$
    \<    \CB  _{P_m}^{(-)} ....\CB  _{P_1}^{(-)} 
    \CB  _{K_1}^{(+)} ...  \CB  _{K_n}^{(+)}
   \>=
   {(-)^{{n(n-1)\over 2}+{m(m-1)\over 2}}\over 
   \prod_{j=1}^{m-1}  
   \sin 2\pi b (P_1+...+P_j  ) \prod_{l=1}^{n-1}  
   \sin {2\pi\over b} ( K_1+...+K_l  )} .
   $$

\appendix{D}{Evaluation of the basic boundary   correlation 
functions  on
the FZZT brane }

 \subsec{Three-point function}

\noindent
For the   three-point function
 \eqn\triptf{   W_{P_1
, P_2 ,P_3}(s ', s  , s'') =
\< ^{s ''}[\CB _  {P_1}^{(-)}]^{s } [ \CB  _{P_2} ^{(+)}] 
 ^{ s' }[\CB  _{P_3} ^{(+)}]^{s''}\>.}
the functional equations \FUE\ and \FUED\  
 read  (with the assumption that  in the first equation
 $P_1<-{1\over 2b}, P_2>{1\over 2b}$
and in the seccond equation
$P_1<-{b\over 2}, P_2>{b\over 2}$)  
\bigskip

  \eqn\FUEtri{ \eqalign{
  & \sin\( 2\pi b  
P_2\)\  W_{P_1
, P_2 ,P_3}(s ', s \pm ib  , s'')   =\cr
& = 2M\cosh\[\pi b\(\frac{s+s'}{2 } \pm iP_2\)\]
  \cosh\[\pi b\(\frac{s-s'}{2 } \pm iP_2\)\] \ 
 W_{P_1+{b\over 2}, P_2-{b\over 2},P_3}(s ', s   , s '') \cr
& + D_{P_1+{b\over 2}}(s, s')
 ;}}

\bigskip
\eqn\FUEDtri{\eqalign{&
  \sin\( \frac{2\pi}{b }  P_2\) \ W_{P_1
, P_2 ,P_3}(s ', s \pm \frac{i}{b}  , s'')   =\cr
&
=2M^{1/b^2} \cosh\[\frac{\pi}{b }  \(\frac{s+s'}{ 2 } \pm iP_2\)\]
  \cosh\[\frac{\pi}{b }  \(\frac{s-s'}{ 2 } \pm iP_2\)\]
  \ W_{P_1+{1\over 2b}, P_2-{1\over 2b},P_3}(s ', s   , s '') \cr
 & + D_{P_1+{1\over 2b}}(s, s').
}
}
Their common solution is given by the boundary three-point function 
in pure   Liouville  theory (with the restriction $P_1+P_2+P_3= e_0/2$  
on the momenta and rescaled  to match with  our definition of the 
boundary cosmological constant), 
which  has  been  found explicitly as an 
integral of products of double sine functions \PTtwo.
 If  the three 
momenta 
are degenerate, that is of the form $( r/b-sb)/2$, the answer 
is  a    rational   function of   trigonometric functions \KPS.

\subsec{Two-point function}

\noindent
The   functional equations
\FUE\ and \FUED\  are derived for the correlation 
functions of three or more operators.
The    two- and one-point  functions in \QG\ are  more subtle quantities
whose correct normalization requires to tame the residual
global conformal symmetry.
 To avoid facing this difficult technical problem, we will 
   apply  a    trick: we will  use  the
  relation between  the two-point function 
  \eqn\twoptf{
D_P(s, s')=  \<^{s'} [\CB  _{-P }^{(-)}]^{ s } 
[  \CB  _{P}^{(+)} ]^{ s '}\>_{\rm disc}
  }
  and the  three-point function
    with  $-P_1=P_2 =P$ and $P_3=\hf e_0$, and 
    then  apply  \FUE\ and \FUED.
 The relation is \foot{The easiest way  we know 
  to prove that is to
  perform a Laplace transformation  
  from the boundary cosmological constant $z(s)$ to the 
  physical length  
  $  \ell =  \int dx e^{ b\phi(x)} $
  for each segment of the boundary.}

\eqn\xxx{
W_{-P, P, \hf e_0}(s ', s   , s '')
=-{D_P(s, s')-D_P(s, s'')\over z(s')-z(s'')}.
}
  In particular, taking $s'=s''$ we have the more familiar relation
  \eqn\tptript{
  {\p D_P(s, s')\over \p z(s')}= - W_{-P 
, P ,\hf e_0}(s ', s  , s'). 
}
In this case  eqn. \FUEDtri\   reads
$$\eqalign{
&\sin\( \frac{2\pi}{b }  P\) [D_P(s \pm ib , s')-D_P(s \pm ib , 
s'')]
=\cr
&
2M  \cosh\[\frac{\pi}{b }  \(\frac{s+s''}{ 2 } \pm iP\)\]
  \cosh\[\frac{\pi}{b }  \(\frac{s-s''}{ 2 } \pm iP\)\]
  [D_{P-\frac{b}{2}}(s, s')-D_{P-\frac{b}{2}}(s, s'')]
  \cr &+ [(z(s')-z(s'')] D_{P-\frac{b}{2}}(s', s''),}
$$
  where it is assumed that  $P>\frac{b}{2}$, so that the shifted 
momentum 
  has the same sign. 
   This equation   implies  the  following linear difference 
  equation  for two-point function:
 \eqn\FUtp{
   {D_{P }(s\pm ib,s')\over D_{P -{b\over 2}  }  (s,s') }
     = {
2M\cosh\[\pi b\(\frac{s+s'}{2 } \pm iP\)\]
  \cosh\[\pi b\(\frac{s -s'}{2 } \pm iP\)\] 
  \over \sin\( 2\pi b  P  \) }.
  }
Similarly one obtains,   for
$P>\frac{1}{2b}$, the dual equation
 \eqn\FUtpD{
 {D_{P} ( s \pm \frac{i}{b}   , s')  \over
D_{P -{1\over 2b} }(s , s')}
  ={
2M^{1/b^2} \cosh\[\frac{\pi}{b }  \(\frac{s+s'}{ 2 } \pm iP\)\]
  \cosh\[\frac{\pi}{b }  \(\frac{s-s'}{ 2 } \pm iP\)\]
 \over \sin\( \frac{2\pi}{b }  P\)   }.
}
    The form of equations  \FUtp\ and \FUtpD\    suggests to
    look for a solution in a factorized form
   \eqn\DDhat{
    D_P(s,s')=\hat D_P(s+s')\hat D_P(s-s')
      }
  where $\hat D_P(s)$ should satisfy
  \eqn\factD{ {\hat D_{P }(s\pm ib)\over
  \hat D_{P-\frac{b}{2}}(s)}
  =2 M^{1/2}\ { \cosh[\pi  b(\frac{s}{2} \pm i P)]\over 
   \sqrt{2\sin\( 2\pi b  P\)  } } \ \qquad (P>\frac{b}{2}).
 }
 and
  \eqn\factDd{ {\hat D_{P }(s\pm\frac{ i}{b})\over
   \hat D_{P-\frac{1}{2b}}(s)}
  =2 M^{1/2b^2} \ {\cosh[\frac{ \pi}{b}(\frac{s}{2} \pm i P)]\over
 \sqrt{2\sin\( 2\frac{\pi}{ b}  P\)  }} \qquad (P>\frac{1}{2b}).
  }
  In addition the two-point function should be  real  even function 
of $P$,
  $s$ and $s'$.
     The     solution is  proportional to the 
     Liouville reflection amplitude
     and is unique up to   normalization:
   \eqn\WPD{
   D_{P}(s_1,s_2)=  { b\over | \sin ( 2\pi  P/b)|}
   D_{|P|}^{(+-)}(s_1,s_2) = 
   {1/ b\over  |\sin (2\pi Pb)|}
     D_{-|P|}^{(-+)}(s_1, s_2) .}
  Here the normalization is fixed  by the requirement   that 
   the amplitude is self-dual.

    \subsec{One-point function}
    
        \noindent
One can  proceed in the similar way to evaluate the 
    boundary one-point function, or the disc loop amplitude
     \eqn\oneptf{
     W(s)= \<[ \CB   _{e_0/2}]^{s,s }\>_{\rm disc}
     }
      using its relation with the 
  boundary two-point function
  \eqn\Weo{ 
  D_{e_0/2}(s_1, s_2)=- {W(s_1) - W(s_2) \over z(s_1) - z(s_2)}.
  }
The explicit expression  of $D_{e_0/2}$ follows from
the last formula in  \exmples:
 \eqn\Weeo{ 
 D_{e_0/2}(s ,s')=-   b   \  {  M^{1/b^2-1} \over  \sin 
(\frac{\pi}{b}e_0)} \
{\cosh( \frac{\pi}{b}s)-\cosh( \frac{\pi}{b}s')\over
\cosh (\pi b s) -\cosh (\pi b s')}.
}
 Comparing \Weeo\  and \Weo\  we get  
  \eqn\Wtau{
 W(s	) =  b  {  M^{1/b^2} \cosh ( \frac{\pi}{b}s)  \over \sin  
(\frac{\pi}{b}e_0)}
 .}
Similarly one obtains the dual loop amplitude
 \eqn\Wdual{
  \tilde W(s) = - {1\over b}   M \
{\cosh (\pi b s)    \over \sin (\pi be_0)}
.
 }
 
 \subsec{The disc partition function of the FZZT brane}

The one-point function is related to the disc partition function 
$S(z)$ by  
\eqn\WvsS{
W(s)= 
 - {\p S(\L, s) \over \p z }, \qquad z(s) = M\cosh\pi b s.
 }%
 Integrating in $z(s)$  gives for the disc partition function
\eqn\partf{
\eqalign{
S(\L, z)&= { b^2 M^{1+1/b^2}\over 2 \sin(\frac{\pi}{b} e_0)}
\( {\cosh(\pi Qs)\over Q}-{\cosh(\pi e_0s)\over e_0}\)\cr
&= {M^{1+1/b^2}\over \sin(\frac{\pi}{b} e_0)}
{b^2\over Qe_0}\(    b \cosh(\pi bs)\cos (  \frac{\pi }{b} s) -
  \frac{1}{b} \sinh(\pi bs)\sinh (  \frac{\pi }{b} s) \).
}
}
     The dual partition function is
     \eqn\partf{
\eqalign{
\tilde S(\L, z)&= -{ b^{-2} M^{1+1/b^2}\over 2 \sin({\pi}{b} e_0)}
\( {\cosh(\pi Qs)\over Q}+{\cosh(\pi e_0s)\over e_0}\)\cr
&= {M^{1+1/b^2}\over \sin({\pi}{b} e_0)}
{b^{-2}\over Qe_0}\(  -\frac{1}{b}  \cosh(\pi bs)\cos (  \frac{\pi 
}{b} s) +
  b\sinh(\pi bs)\sinh (  \frac{\pi }{b} s) \)
}
}
The two partition functions are related by a Legendre transformation:
\eqn\Legtr{
\frac{1}{b}\sin(\frac{\pi}{b} e_0)  S(s) +  b    \sin({\pi}{b} e_0)\  
\tilde S(s)
+ \tilde z(s) z(s)=0
.}


\appendix{E}{Bilinear equation for the loop amplitude}


          \noindent
         The   Virasoro conditions on the  loop amplitude  \defWxd\ 
         are 
           \eqn\lopex{
         \tilde W_x^2( \tilde z) +   \oint _{\CC_1}
       { \tilde z' \over \tilde z} 
        \ {  \tilde W_x( \tilde z)-   \tilde W_x( \tilde z')\over  \tilde z- \tilde z'} 
        \[2T +   \tilde W_{x+1}(2T- \tilde z')
        +   \tilde W_{x-1}(2T- \tilde z')\]{d \tilde z'\over 2\pi i} =0
}
where the contour  $\CC_1$ encircles the  support of the 
eigenvalue density   $[0,a]$
  but not the
interval $[2T-a, 2T]$.   For   amplitudes of the form    \Wxxd\
this  equation becomes 
  \eqn\lope{
         \tilde W^2( \tilde z) +   \oint _{\CC_1}
       { \tilde z' \over \tilde z} 
        \ {  \tilde W( \tilde z)-   \tilde W( \tilde z')\over  \tilde z- \tilde z'} 
        \[2T +2\cos  (\pi b^2) \  \tilde W(2T- \tilde z')\]{d \tilde z'\over 2\pi i} =0.
}
 The function $\tilde W(\tilde z)$  is   meromorphic 
 with a cut along the support  $[0, \tilde a]$ of the eigenvalue distribution.
   Equation  \lope\  implies   
       the boundary condition along the cut
          \eqn\lpqzd{
     W(\tilde z+i0) +W(\tilde z-i0)-  2\cos  (\pi b^2)\
      W (2T - \tilde z)
     =2T \qquad (  0<\tilde z< \tilde a) .
     }

The integral equation  \lope\ can be turned into a functional equation 
by the following trick ( \Idis,  sect. 3.2).  Replacing  $ \tilde z$ by $2T- \tilde z$ in
\lope\ and adding  the two equations  one  can  turn  the integral
along the contour $C_1$  into an integral along a contour $\CC_{\infty}$,
 which encircles  both intervals
$[0,a]$ and $[2T-a, 2T]$ and therefore can be expanded to infinity.
 Now the   integration  can be performed by the residue formula
  using the asymptotics 
     \eqn\asymWz{
      \tilde W( \tilde z)= {N/  \tilde z} +\tilde W_1/\tilde z^2+ ...\qquad (\tilde z\to\infty)
     }
     which  yields the 
     the functional identity:
    \eqn\LOOPEQqq{\eqalign{
     &   \tilde W^2( \tilde z) +  \tilde W^2(2T- \tilde z)  
     -2 \cos (\pi b^2)   W ( \tilde z) W( 2T- \tilde z)
     \cr &
      =2 T\[   \tilde W( \tilde z) +  W( 2T- \tilde z)\]
     - {4T^2 \tilde C\over \tilde z(2T-\tilde z)} .}}                
   Here the constant $\tilde C =\tilde C(N,T)$ is given by the   integral    
   \eqn\pocham{
          \tilde C=     {2\cos (\pi b^2)\over T} \  \oint _{\CC_1}
      {d \tilde z'      \over 2\pi i}   W( \tilde z') \tilde W(2T- \tilde z').
}

        The critical value of $\tilde C$ can be found by  solving \LOOPEQqq\
     for $\tilde z=T$, where  it  becomes algebraic. This gives
\eqn\Wofo{ \tilde W(T  )=
 {  T \over 2 \sin^2( \pi b^2)}-{\sqrt{\tilde C_c-  \tilde C}\over 
 \sin( \pi b^2) }
      ,\quad
 \tilde C_c = { T^2\over  4 \sin^2( \pi b^2)   }.}     
 At the critical point the solution of       \LOOPEQqq\
 is given by
   \eqn\zofu{ \tilde z-T =  {T\over \cosh  u}, \quad
   \tilde W =  A+ B {\sinh  (1-b^2)u \over \sinh   u},
   }
 with              $A= {T \over 2\sin^2 (\pi b^2/2)}$ and $
       B= {T \over \sin (\pi b^2/2)  \sin (\pi b^2) }$.
The coefficient $B$ is such that   
$\tilde W(\tilde z)=0$  at  $\tilde z=\infty$.
 For given $T$ the critical value of $N$ is determined by 
 comparing   the large $\tilde z$ asymptotics of the solution 
 \zofu\ with   \asymWz: 
\eqn\critt{  N_c =  { 1-b^2 \over \sin ({\pi b^2})} \ T^2
.}
     For  $\L\equiv N-N_c<<N_c$   
     the constant $\tilde C$  behaves as
               $$ \tilde C = \tilde C_c +c_1 \L + c_2 \L^{1/b^2}  T^{2-2/b^2}  + ...$$
    and   as  $b^2<1$, only  
     the   first term  survives in the scaling limit $\L/T^2\to 0$.

\listrefs

\bye